\documentclass{Liquid_film}

\usepackage{graphicx}
\usepackage{natbib}
\usepackage{subfig}
\usepackage{amsmath}
\usepackage{amssymb}
\usepackage{color}
\usepackage{wasysym}

\title{Falling liquid films with blowing and suction}

\author[A. B. Thompson, D. Tseluiko and D. T. Papageorgiou]
{Alice B. Thompson$^1$\thanks{Email address for correspondence: alice.thompson1@imperial.ac.uk}, Dmitri Tseluiko$^2$ and Demetrios T. Papageorgiou$^1$}

\affiliation{$^1$ Department of Mathematics, Imperial College London, London, SW7 2AZ, UK.\\[\affilskip]
$^2$Department of Mathematical Sciences, Loughborough University, Loughborough LE11 3TU, UK.}

\pubyear{2010}
\volume{650}
\pagerange{119--126}
\date{}
\begin{document}
\maketitle

\begin{abstract} 
Flow of a thin viscous film down a flat inclined plane becomes unstable to long wave interfacial fluctuations when the Reynolds number based on the mean film thickness becomes larger than a critical value (this value decreases as the angle of inclination with the horizontal increases, and in particular becomes zero when the plate is vertical). Control of these interfacial instabilities is relevant to a wide range of industrial applications including coating processes
and heat or mass transfer systems. This study considers the effect of blowing and suction through the substrate in order to construct from first principles physically realistic models that can be used for detailed passive and active control studies of direct relevance to possible experiments. Two different long-wave, thin-film equations are derived to describe this system; these include the imposed blowing/suction  as well as inertia, surface tension, gravity and viscosity. The case of spatially periodic blowing and suction is considered in detail and the bifurcation structure of forced steady states is explored numerically to predict that steady states cease to exist for sufficiently large suction speeds since the film locally thins to zero thickness giving way to dry patches on the substrate. The linear stability of the resulting nonuniform steady states is investigated for perturbations of arbitrary wavelengths, and any instabilities are followed into the fully nonlinear regime using time-dependent computations. The case of small amplitude blowing/suction is studied analytically both for steady states and their stability. Finally, the transition between travelling waves and non-uniform steady states is explored as the suction amplitude increases.
\end{abstract}

% \begin{keywords}
% % Thin films, long wave equations, bifurcation
% \end{keywords}

\section{Introduction}

The flow of a viscous liquid film down an inclined plane, under the action of gravity, inertia and surface tension, is a fundamental 
problem in fluid mechanics that has received considerable attention both theoretically
and experimentally due to the richness of its dynamics and its wide technological applications, e.g. in coating processes and heat or
mass transfer enhancement. 
For sufficiently thin films (with other parameters such as inclination angle and viscosity fixed), 
the uniform Nusselt solution \citep{Nusselt} is stable, but for thicker layers the flow is susceptible to interfacial instabilities in the form of 
two-dimensional travelling waves which propagate down the slope, followed by more complicated time-dependent and three-dimensional behaviour.
The linear stability of a uniform film was first considered by \citet{Benjamin} and \citet{Yih}, who used an Orr--Sommerfeld analysis to show that instability first appears at wavelengths that are large compared to the undisturbed film thickness $h_s$. Using the Nusselt velocity at the free surface we define a Reynolds number $R=\rho^2 g h_s^3\sin\theta/2\mu^2$ (see \eqref{eq:ReyCap} also) where
$\rho$ is the fluid density, $g$ is the gravitational acceleration, $\theta$ is the angle of inclination to the horizontal (see Figure \ref{fig:Flow_diagram}), and $\mu$ is the viscosity of the fluid; the flow becomes linearly unstable to long waves when $R>R_c=
(5/4)\cot\theta$ and we can see
that the critical Reynolds number tends to zero as the plate becomes vertical. This result also shows that for a given angle
the flow can be made unstable by increasing the density and/or the film thickness, or decreasing the viscosity.

In order to go beyond linear theory without recourse to direct numerical simulations of the Navier-Stokes equations,
a hierarchy of long-wave reduced-dimension models have been developed to analyse in detail the stability and nonlinear 
development of the flow \citep[see reviews by][]{Craster_review,Kalliadasis}.
The simplest fully nonlinear long-wave model was developed by \citet{Benney} who used an expansion in a small slenderness parameter to 
derive a single evolution equation for the interface height. 
The Benney equation is valid for Reynolds numbers near the critical value $R_c$
(in fact it captures exactly the linear stability threshold), in the sense that far
away from critical in the unstable regime solutions can become unbounded in finite time \citep{Pumir}, a phenomenon that invalidates the
long wave approximation and is not observed in numerical simulations of the Navier-Stokes equations \citep{ Oron_Gottlieb,Scheid_et_al}.
The Benney equation forms a rational basis for the development of asymptotically correct weakly nonlinear models that lead
to the Kuramoto-Sivashinsky (KS) equation \citep[see][and references therein, for example]{Waves_on_electrified_thin_films}. 
The KS equation displays very rich dynamical behaviour including spatiotemporal chaos. In other canonical asymptotic weakly nonlinear regimes one can derive the generalised (i.e. dispersively modified)
KS equation along with electric-field induced instabilities \citep{Waves_on_electrified_thin_films,Tseluiko-PRE}.
In order to overcome the near-critical restrictions of the Benney equation, 
\citet{Shkadov} developed a coupled fully nonlinear long-wave system for the free-surface height and the local mass flux. 
The Shkadov model avoids finite-time singularities, but underpredicts the critical Reynolds number.
Using a weighted-residual method, \citet{Ruyer_Quil} recently developed a new two-degree of freedom long-wave model, which has the correct stability threshold and is well behaved in the nonlinear regime. In this paper we consider two such reduced-dimensional models in the presence of blowing and
suction.

In applications it is useful to be able to control the film dynamics. For instance in
coating applications a stable uniform film is required, whereas in heat or mass transfer applications efficiency is improved if the
flow is nonuniform and attains increased surface area and recirculation regions. 
Such diverse requirements motivate the introduction of extrinsic modifications to the system in the interest of
controlling the dynamics. An example of such a modification is
the utilization of a heated substrate which affects the interfacial dynamics via a combination of Marangoni effects and evaporation \citep{Kalliadasis}. Substrate heating thus introduces new modes of instability relating to convection, and lowers the critical Reynolds number. The substrate behaviour can also be altered by allowing chemical coatings, elastic deformations, or interactions with flow through porous media \citep{Thiele, Ogden_etal, Samanta_et_al_2011, Samanta_et_al_2013}, which is often modelled by an effective slip condition. 
External fields can also be used to stabilise or destabilise the interface. Depending on the fluids used,
an applied magnetic field can stabilise
the flow \citep[see][]{Hsieh1965, Shen_magnetic, RenardySun}, whereas an electric field applied normal to the interface
can destabilize the flow and in fact drive it to chaotic spatiotemporal dynamics even
below critical $R<R_c$ \citep{Waves_on_electrified_thin_films,Tseluiko-PRE}.

One way to modify the dynamics of falling film flows is to topographically structure the substrate. There have been several theoretical and
experimental studies of film flows down wavy inclined planes (typically with sinusoidal and step topographies), aiming to explore how
topography affects stability and stability criteria such as critical Reynolds numbers, how substrate heterogeneity interacts with
nonlinear coherent structures, and from a practical perspective how topography induces flows that can be useful
in heat or mass transfer by creating regions of recirculating fluid \citep[see for example][and numerous references therein]{Inclined_topography}. The problem is quite complex with several parameters and an overall conclusion of these studies is that topography can
either decrease or increase the critical Reynolds number in different regimes.

Inhomogeneous heating of the substrate can generate nonuniform film profiles even in the absence of inclination \citep{Saprykin}.
\citet{Nonuniform_heating} used an extension of the Benney model to analyse flow over a substrate with a sinusoidally varying heat distribution, where temperature is coupled to flow via Marangoni effects. They solved the equations to obtain travelling waves in the case of uniform heating, and steady non-uniform solutions in the case of non-uniform heating, and used initial value calculations to demonstrate that imposing heating is able to halt the progress of a travelling wave, leading to stable steady interface shapes.
The combination of localized heating and topography has also been considered and it has been demonstrated that features
that form in the isothermal case (e.g. ridge formation ahead of step-down topography) can be removed by suitable heating \citep[see][]{Blyth_Bassom, Ogden_etal}.
The removal of such a ridge has also been shown to be possible by the imposition
of vertical electric fields rather than heating, providing another physical mechanism for interface control \citep[see][]{Tseluiko_et_al_2008b,Tseluiko_et_al_2008a}.

Suction and injection blowing through an otherwise rigid substrate has well known applications in stabilizing
flows and changing global structures that can negatively affect performance, such as boundary layer separation for example.
In the types of interfacial flows of interest here there has been much more limited exploration;
\citet{Momoniat2010} studied the effect of imposing either suction or injection on a spreading drop and found that injection enhances ridge formation, while suction leads to cavities forming on the free surface. As the total mass is not conserved, a steady state is impossible, but they found that both injection and suction are able to break up streamlines.
The total fluid mass is also not conserved for the flows of films and drops over porous substrates \citep{DavisHocking}; in fact both drops and films are drawn entirely into the substrate in finite time as would be expected. 
In the analysis of drop evaporation, a number of studies, e.g. \cite{AndersonDavis} and more
recently \citet{Todorova_et_al_2012} among others, have considered the steady state obtained by imposing injection through the substrate that exactly balances the mass lost to evaporation. 
In the latter study the injection profile was imposed according to a Gaussian distribution, and the drop shape is largely independent of this distribution as long as the injection is not too large near the drop contact line
 \citep[note that a precursor film model was used by][]{Todorova_et_al_2012}.
 For continuous falling liquid films over porous substrates there have been linear stability studies invoking
a Darcy law in the porous medium and a Beavers-Joseph boundary condition at the liquid substrate boundary
\citep{SadiqUsha2008, Usha_et_al_2011}. It is found that an effective slippage takes place that enhances the
instability in the sense that it reduces the critical Reynolds number. Slippage models were investigated further
by \citet{Samanta_et_al_2011} and an alternative porous medium model is proposed and analysed by \citet{Samanta_et_al_2013}.

In this paper, we impose blowing or suction through the wall, and perpendicular to it, of fluid that is identical
to that of the liquid film. The magnitude of the blowing/suction is assumed to vary spatially along the planar substrate and
hence modifies the no penetration boundary condition there, but we assume that there is no slip along the substrate. We envision that such a model would be appropriate to
experimental setups where tiny slits on the substrate would provide the conduit for fluid to enter and leave the wall.
Such mechanisms affect the total mass in the film and on physical grounds we can anticipate that a net suction would
dry the substrate in finite time whereas a net blowing increase the total mass and hence the mean thickness at
any given time. An increase in thickness would consequently increase the local (in time) Reynolds number since it is proportional to the mean thickness,
hence the flow is expected to become more unstable. In this study we will consider the case of blowing/suction that conserves
the total film mass (e.g. spatially periodic blowing/suction of zero mean), which is possibly the most interesting case since it sits
on the boundary of the net mass decrease or increase, and hence both stabilising and destabilising phenomena can occur depending on the parameters,
as will be seen later.

The rest of the paper is organised as follows. In \S \ref{sec:Formulation}, we discuss the governing equations and dimensionless parameters, the scaling and statement of the two long wave models, and the choice of the blowing and suction function. The numerical methods used to solve these models are described in \S \ref{sec:Numerical_methods}.
In \S \ref{sec:Steady_bifurcation_structure}, we explore the steady states and bifurcations obtained for non-zero imposed suction, discovering a non-trivial bifurcation structure even at zero Reynolds number. We also discuss the distinctive effect of the suction on flow streamlines. 
Linear stability of steady solutions is discussed in \S \ref{sec:Linear_stability},
 with a focus on stability to perturbations of arbitrary wavelength, and thus the effect of suction on the critical Reynolds number. In \S \ref{sec:Travelling_waves}, we investigate the effect of imposing suction on the travelling waves which occur above the critical Reynolds number. In \S \ref{sec:IVP}, we review the various initial value calculations performed in this paper, and provide further results. Finally, we present our conclusions in \S \ref{sec:Conclusion}.

\section{Problem formulation} 
\label{sec:Formulation}
\subsection{Nondimensionalisation and scaling}

\begin{figure}
\centering
\includegraphics{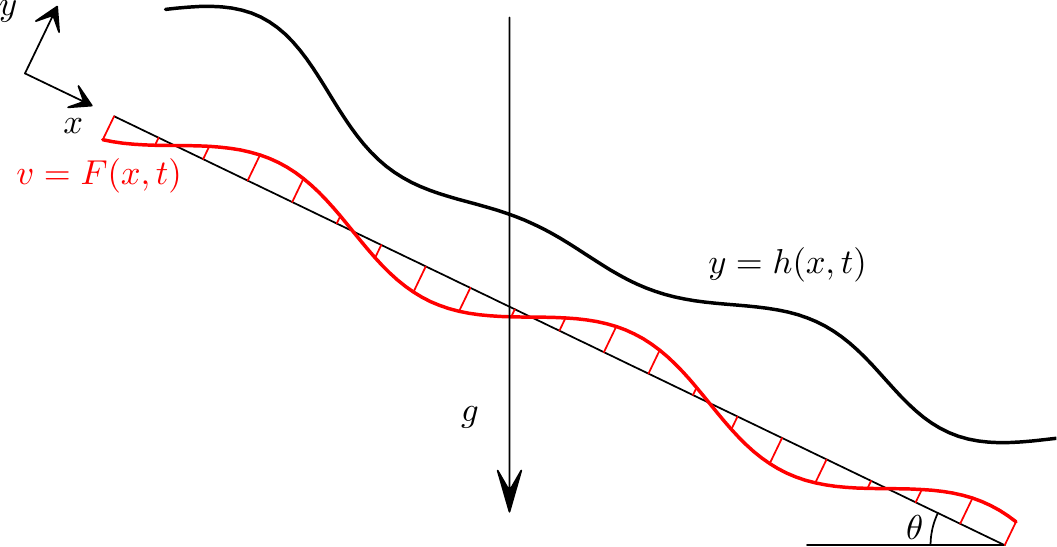}
 \caption{Sketch of flow domain showing coordinate system. We consider a fluid layer, with mean height $h_s$, bounded along $y=0$ by a rigid wall inclined at angle $\theta$ to the horizontal, and at $y=h(x,t)$ by a free surface. Fluid is injected through the wall, and so the normal velocity at the wall is given by the prescribed function $v = F(x,t)$.
 \label{fig:Flow_diagram}}
\end{figure}

We wish to determine the evolution of a falling liquid film, with mean thickness $h_s$, on a slope inclined at angle $\theta$ to the horizontal. We model the liquid as a Newtonian fluid of constant dynamic viscosity $\mu$ and density $\rho$, and the air as a hydrodynamically passive region of constant pressure $p_a$. The coefficient of surface tension across the air/liquid interface is $\gamma$. We assume that the film flow is two-dimensional, with no variations in the cross-stream direction.
We use coordinates as illustrated in figure~\ref{fig:Flow_diagram}; $x$ is the down-slope coordinate, and $y$ is the coordinate in the direction perpendicular to the slope, so that the wall is located at $y=0$ and the interface is defined as $y=h(x,t)$. 
We denote the velocity components in the $x$ and $y$ directions as $u$ and $v$ respectively.

The dimensionless film flow is governed by the Navier--Stokes equations 
 \begin{equation}
 \label{non-scaled-start}
R\left(u_t + u u_x + vu_y\right) 
= -  p_x
 +2+u_{xx}+ u_{yy},
\end{equation}
 \begin{equation}
R \left(v_t + u v_x + v v_y\right) =
-  p_y
-2 \cot\theta 
+v_{xx}+v_{yy},
\end{equation}
which are coupled to the incompressibility condition
 \begin{equation}
 u_x + v_y=0.
\end{equation}
Here we have non-dimensionalised the equations using the mean film thickness $h_s$ as the length scale, the Nusselt surface speed of a flat film $U_s=\rho g h_s^2\sin\theta/2\mu$ \cite[][]{Nusselt} as the velocity scale, $h_s/U_s$ as the time scale, and $\mu U_s/h_s$ as the pressure scale. We note that this scaling, based on surface speed, is the same scaling as used by \citet{Inclined_topography} to study the influence of wall topography on the stability of flow down an inclined plane.
We define the Reynolds number $R$ and the capillary number $C$ based on the surface speed $U_s$, so that
\begin{equation}
 R = \frac{\rho h_s U_s}{\mu}, \quad C = \frac{\mu U_s}{\gamma}.\label{eq:ReyCap}
\end{equation}

We will suppose that the imposition of suction boundary conditions at the wall does not alter the no-slip condition, but does affect the no-penetration condition, so that the complete boundary conditions at the wall, $y=0$, are
\begin{equation}
 u =0, \quad  v=F(x,t).
\end{equation}
At the interface, $y=h(x,t)$, the tangential and normal components of the dynamic stress balance condition yield
\begin{eqnarray}
&\left (v_x+u_y\right)\left (1-h_x^2\right) + 2h_x \left(v_y-u_x\right)=0,&\\
&\displaystyle  p - p_a
 - \frac{2}{1+ h_x^2} 
 \left(
v_y
+ u_x h_x^2   - h_x\left(v_x+u_y\right)\right) 
 = -\frac{1}{C}\frac{h_{xx}}
 {
 (1+h_x^2)^{3/2}
 },&
 \label{non-scaled-end}
\end{eqnarray}
respectively.
The kinematic boundary condition on the interface can be written in integral form as 
\begin{equation}
\label{mass_governing}
 h_t - F(x,t) + q_x=0, 
\end{equation}
where $q(x,t)$ is the stream-wise flow rate, defined as
\begin{equation}
 q(x,t) = \int_0^h u(x,y,t) \, \mathrm{d}y.
\end{equation}

\subsection{Long-wave evolution equations}

We now seek solutions with wavelength $L\gg 1$, and we introduce the long-wave parameter $\delta = 1/L$. We derive two first-order long-wave models, based on an asymptotic expansion in the long-wave parameter \cite[a Benney-type model, see][]{Benney} and on a weighted-residual method \cite[following the approach of][]{Ruyer_Quil}; derivations of both models are presented in appendix~\ref{sec:Derivation_appendix}. We assume that $\cot \theta=O(1)$. To retain both inertia and surface-tension effects, we additionally assume that $R=O(1)$ and $C = O(\delta^2)$. We choose the canonical scaling $F=O(\delta)$ so that the imposed suction can enter and compete with the perturbed flow and hence facilitate possible instability enhancement or reduction.

The essential task of the derivation is to estimate the flow rate  $q(x,t)$ for a non-uniform film.
In both models, the mass conservation equation is unchanged from \eqref{mass_governing}, and so we have
\begin{equation}
\label{eq:mass}
 h_t - F + q_x=0.
\end{equation}
However, the two models differ in their treatment of nonlinearities in the momentum equation, and thus yield different equations for $q$.

In the Benney equation (see \S~\ref{sec:Benney_derivation}), $q$ is slaved to the interface height $h$, and is given by 
\begin{equation}
\label{eq:Benney}
 q(x,t) =  \frac{2h^3}{3} 
 -\frac{h^3  }{3}\left(2 h\cot\theta - \frac{h_{xx}}{C}\right)_x
+
R \left( 
\frac{8h^6h_x }{15}
-\frac{ 2h^4 F}{3}
\right).
\end{equation}
The first-order weighted-residual approach (see \S~\ref{sec:WR_derivation}) instead yields an evolution equation for $q$:
\begin{equation}
\label{eq:WR}
 \frac{2}{5}R h^2 q_t + q = \frac{2h^3}{3}-\frac{h^3}{3}\left(2 h \cot\theta - \frac{h_{xx}}{C}\right)_x 
 + R \left( \frac{18q^2 h_x}{35} - \frac{34 h q q_x}{35}+ \frac{hqF}{5}\right).
\end{equation}
The Benney and weighted-residual equations are identical when $R=0$.
For $R\neq0$, the Benney equation has a single degree of freedom $h(x,t)$, while the weighted-residual model has two degrees of freedom, $h(x,t)$ and $q(x,t)$. As a result, the weighted-residual equations can in principle exhibit richer behaviour. However, despite the additional complexity, the weighted residual equations in fact support bounded solutions across a greater range of parameter space \citep{Scheid_et_al}.

\subsection{Choice of blowing and suction function}
\label{sec:Form_of_forcing}

In time-dependent evolution, the mean layer height is conserved only if the imposed flux function $F$ has zero mean, and hence steady states can only exist if $F$ has zero mean.
Conserved mean layer height is the natural state for numerical calculations in a periodic domain, and is sometimes called `closed' conditions, as there is no net flux out of the domain. However, in experiments, closed conditions are not easy to implement, and so experimental realisations more typically impose the fluid flux at the domain inlet. In this case, there is no direct control over the mean layer height.

In the absence of blowing and suction, both the open and closed systems support uniform flow via the Nusselt solution, and thus we obtain the same scaling whether based on the mean layer height or mean flux.
 In order to investigate the influence of suction, we can consider steady states where the mean layer height remains fixed for closed conditions, or where there is no change to the mean flux for open conditions.
However, the bifurcation diagrams for fixed mean layer height correspond most naturally to statements about time evolution, as the mean layer height does not change with time.
Most of the results we present are for fixed mean layer height, but we will also present some results for fixed flow rate, and we note that there is a mapping between results for these two conditions.

 In the rest of this manuscript, we will consider the simplest functional form for $F$ with zero mean, i.e. a single harmonic mode:
\begin{equation}
 F(x) = A \cos(mx) = A \cos\left(\frac{2\pi x}{L}\right).
\end{equation}
This function $F(x)$ has the symmetry that the transformation $A\rightarrow -A$ is recovered by translation in $x$ by a distance $L/2$, and so translationally-invariant solution measures, such as the critical Reynolds number for onset of instability, must be even in $A$.

The long wave equations \eqref{eq:mass}, \eqref{eq:Benney} and \eqref{eq:WR} also apply if $F$ is unsteady, so long as $F$ varies on dimensionless timescales no shorter than $O(1/\delta)$, or the dimensional timescale $h_s/(\delta U_s)$.
Time derivatives of the vertical velocity, and hence $F$, would feature in the equations at the next order in $\delta$. We are therefore free, at this order, to impose a time-dependence on $F$, or even choose $F$ in response to the film evolution. The latter formulation would be particularly useful in feedback control studies.

\section{Numerical methods}
\label{sec:Numerical_methods}

The numerical calculations that we perform are of three types: computation of steady periodic solutions and their bifurcation structure, linear stability calculations of such steady states to perturbations of arbitrary wavelength, and nonlinear time-evolution via initial value problems. We conduct these calculations using the continuation software package \textsc{Auto-07p} \citep{Auto_07p} and Matlab.

The first task, of computing steady solutions, and exploring their bifurcation structure, was conducted in \textsc{Auto-07p} by formulating the problem as a boundary value problem with periodic boundary conditions. 
The \textsc{Auto-07p} code is spatially adaptive, and unlikely to return spurious solutions.  
We are particularly interested in limit point, pitchfork and Hopf bifurcations. The first two of these correspond to bifurcations of steady states, and so can be detected and tracked using the same formulation as for standard steady states.
With regard to Hopf bifurcations, here we are concerned with Hopf bifurcations with respect to time, whereby a steady state becomes oscillatory in time when subject to a perturbation of fixed spatial wavelength. If the perturbation satisfies periodic spatial boundary conditions, the instability will in fact be oscillatory in both space and time. 
We used \textsc{Auto-07p} to track individual Hopf bifurcations by manually augmenting the steady system with a boundary value problem for the spatially periodic but non-constant eigenfunction, with the temporal eigenvalue determined as part of the solution.

The linear stability calculations were performed in Matlab, and we used a pseudo-spectral method for the spatial discretisation.
After spatial discretisation, the governing partial differential equation becomes a large system of coupled first-order ordinary differential equations, which we can write in the general form
\begin{equation}
\label{general_residual}
 \mathbf{F}(\mathbf{u}, \mathbf{\dot {u}})=0
\end{equation}
so that steady solutions $\mathbf{u}_0$ satisfy $\mathbf{F}(\mathbf{u_0}, \mathbf{0})=0$.
In order to determine the linear stability of a steady solution, we suppose that
\begin{equation}
 \mathbf{u}(t) = \mathbf{u}_0 + \epsilon\mathbf{v}\exp(\lambda t), \quad \epsilon \ll 1.
\end{equation}
We now expand \eqref{general_residual} for small $\epsilon$, to obtain 
\begin{equation}
\label{eigenproblem}
 \mathbf{J} \mathbf{v} + \lambda \mathbf{M}\mathbf{v}=0, \quad \mathbf{J}= \left.\frac{\partial \mathbf{F}}{\partial \mathbf{u}}\right|_{\mathbf{u_0}, \mathbf{0}}, \quad \mathbf{M} = \left.\frac{\partial  \mathbf{F}}{\partial \mathbf{\dot{u}}}\right|_{\mathbf{u_0}, \mathbf{0}},
\end{equation}
which is a generalised eigenvalue problem for $\lambda$ and $\mathbf{v}$,
where $J$ is the Jacobian matrix and $M$ is the mass matrix.
As very few points were needed for the spatial discretisation (we typically used 99 equally spaced points), we solved the eigenvalue problem \eqref{eigenproblem} directly in Matlab. We used Floquet multipliers to determine linear stability to perturbations of arbitrary wavelength, and so modified the Jacobian matrix to account for these when necessary.

We note that in the Benney equations, the only time derivatives are those of interface height $h$, while the weighted-residual equations also feature time derivatives of flux $q$. This means that for the same spatial discretisation, there are twice as many eigenmodes for the weighted-residual equations as for the Benney calculations. We found that the weighted residual calculations were prone to spurious eigenmodes, which we removed by careful comparison of the eigenvalue spectrum at different spatial resolutions. We also neglected the neutrally-stable eigenmode corresponding to increasing the total volume of fluid in the domain, which arises in both sets of equations. 

Time evolution calculations were always performed in a fixed spatially-periodic domain. The spatial problem was discretised via a pseudo-spectral method, while time-derivatives were handled via a second-order backward finite difference scheme (BDF2). The resulting code is fully implicit, and solved via direct Newton iteration.

 The code was verified by comparing the steady solutions and bifurcation structure obtained in Matlab to those obtained in \textsc{Auto-07p}. Further validation was obtained by comparison of numerical results to analytical solutions for the shape of small-amplitude steady states and to analytical results for the linear stability of uniform and small-amplitude states.

\section{Bifurcation structure for steady states}
\label{sec:Steady_bifurcation_structure}

In the absence of blowing or suction, the only spatially-periodic steady state is a uniform film. Introducing periodic suction naturally imposes a spatial structure on the solutions, and means that any steady states must be non-uniform. When $R>0$ the solutions and bifurcations differ between the two long-wave models. 
The weighted residual model avoids the blow-up behaviour sometimes exhibited by the Benney equations, and more accurately represents the behaviour of the Navier--Stokes equations at moderate Reynolds number. 
We will generally present results for the weighted-residual model when the two models differ, but we note that a non-trivial bifurcation structure emerges even at zero Reynolds number.

\subsection{Steady solutions at small $A$}
\label{sec:Small_A}

\begin{figure}
\includegraphics[width=\textwidth]{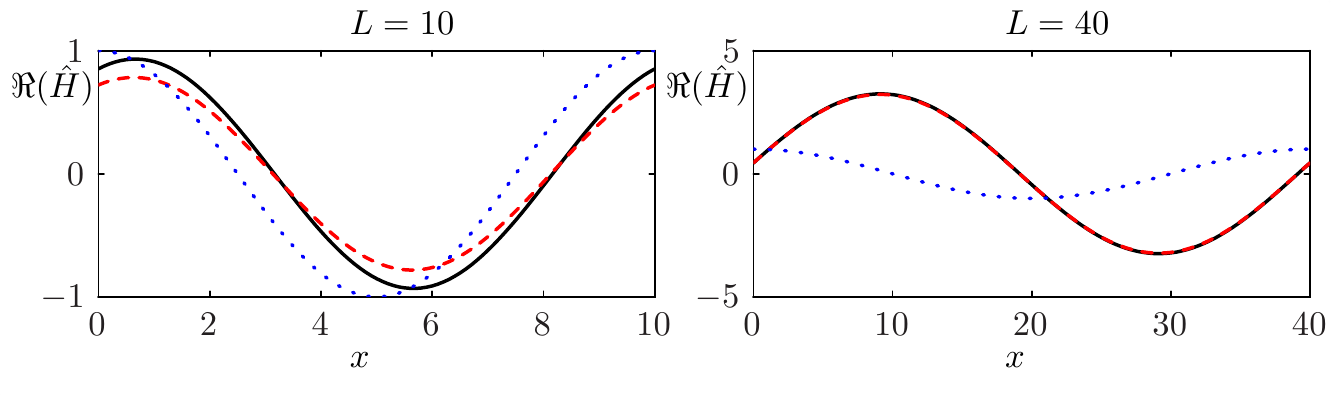}
\caption{The effect of small $A$ on steady solutions with $F=A\cos(2\pi x/L)$ for $L=10$ and $L=40$. The $O(A)$ correction $\Re(\hat{H})$ is shown for the Benney equations (solid line) and the weighted-residual equations (dashed line). The dotted line indicates $F(x)/A$.
In both cases, $C=0.05$, $\theta=\pi/4$ and $R =2$.
\label{fig:Small_A_solutions}
}
\end{figure}

We begin by considering the effect of small-amplitude forcing, in the form of blowing and suction, on the uniform steady state $h=1$.
We choose $F=A \cos{mx}$ where $m=2\pi/L$, and seek a steady solution for $h$ and $q$ when $|A|\ll 1$ of the form
\begin{equation}
 h = 1+ A\Re( \hat{H}) + O(A^2), \quad q = \frac{2}{3}+ A\Re( \hat{Q}) + O(A^2).
\end{equation}
The mass conservation equation \eqref{eq:mass} immediately supplies $\hat{Q} = \exp(imx)/(im)$.
However, the complex constant $\hat{H}$ differs between the two long-wave models.
The Benney result, obtained by linearising \eqref{eq:Benney}, is
\begin{equation}
\label{Small_A_Benney}
 \hat{H}_{\mathrm{Benney}} = \frac{1 + \dfrac{2}{3}imR}{2im +\dfrac{2}{3}m^2 \cot\theta - \dfrac{8}{15} Rm^2 +\dfrac{m^4}{3C}} \exp(imx),
\end{equation}
while the weighted residual equation \eqref{eq:WR} gives
\begin{equation}
\label{Small_A_WR}
 \hat{H}_{\mathrm{WR}} = \frac{1 + \dfrac{18}{35}imR}{2im +\dfrac{2}{3}m^2 \cot\theta - \dfrac{8}{35} Rm^2 +\dfrac{m^4}{3C}}\exp(imx).
\end{equation}
The resulting solutions for $\hat{H}$ are illustrated in figure~\ref{fig:Small_A_solutions} for $L=10$ and $L=40$.

The two expressions \eqref{Small_A_Benney} and \eqref{Small_A_WR} are equal only if $R=0$; otherwise both the magnitude and phase of $\hat{H}$ differ between the two equations.  At $L=10$, the predictions obtained via the two models are in reasonable agreement, but they are indistinguishable when $L=40$. 
Returning to the variables of the long-wave derivation, we set $m = \delta M$ and $C = \delta^2 \widehat{C}$, and expand for small $\delta$; we find that both models yield
\begin{equation}
\label{Long_wave_small_A}
\hat{H} = \frac{1}{2\delta i M}+\left( \frac{\cot\theta}{6}   + \frac{ M^2}{12\widehat{C}} + \frac{R}{5}    \right) + O(\delta),
\end{equation}
which is the expected order of agreement given that terms beyond the second order in $\delta$ were neglected in the derivation of each model. We note from \eqref{Long_wave_small_A} that the magnitude of $\hat{H}$ is inversely proportional to $M$ at leading order, so that for fixed $A$, the maximum perturbation to the interface grows linearly with the wavelength $L$ of the imposed blowing and suction. The long wave expression \eqref{Long_wave_small_A} also reveals a phase shift between $h$ and $F$, which tends to $\pi/2$ as $\delta\rightarrow 0$.

In order to calculate the small $A$ correction to the mean flux analytically, we must expand the steady solution $h =H(x)$, $q=Q(x)$ to $O(A^2)$.
For steady states, with $F(x) A \cos{mx}$, we integrate the mass conservation equation \eqref{eq:mass} to yield
\begin{equation}
\label{Q_steady_expansion}
 Q(x) = \frac{2}{3} + \frac{A \sin{mx}}{m} + Q_2 A^2 + O(A^4)
 \end{equation}
 where we have used the fact that the spatially-averaged flux is even in $A$.
We then expand the interface height in $A$ as
\begin{equation}
\label{H_steady_expansion}
 H(x) = 1 + A(p_1 \cos{mx} + p_2\sin{mx}) + A^2(p_3 \cos{2 mx} + p_4 \sin{2mx}) + O(A^3).
\end{equation}
We solve the flux equation, either \eqref{eq:Benney} or \eqref{eq:WR}, at $O(A)$ and $O(A^2)$ to determine the constants $p_1$, $p_2$, $p_3$, $p_4$ and $Q_2$.
The resulting coefficients are somewhat lengthy and so we do not list them here. In the long wave limit $m = \delta M$,  $C = \delta^2 \widehat{C}$, with $\delta \ll 1$, both the Benney and weighted residual equations yield
\begin{equation}
\frac{1}{L}\int_0^L Q(x) \, \mathrm{d}x = \frac{2}{3} + \frac{A^2}{\delta^2}\left( \frac{1}{4 M^2} + O(\delta^2)\right) + O(A^4),
\end{equation}
so that small amplitude, long wave blowing and suction always increases the mean down-slope flux.

\subsection{Steady solutions as $A$ increases}
\label{sec:Steady_increasing_A}

\begin{figure}
\includegraphics{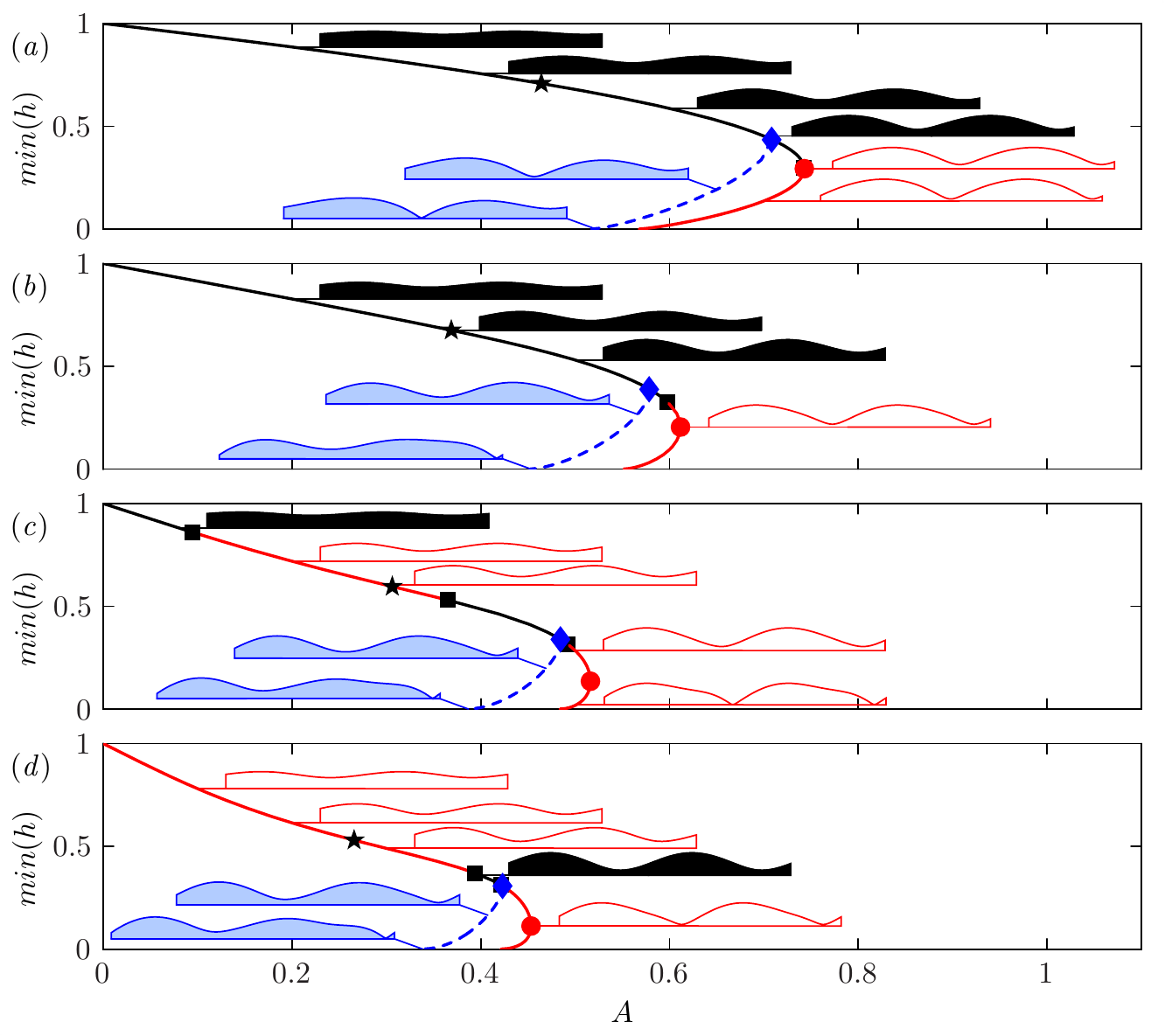}
 \caption{Bifurcation structure for steady solutions to the weighted residual equations as $A$ increases, subject to fixed mean layer height of 1, and calculated in a periodic domain of length $2L$. Here $F=A\cos(2\pi x/L)$, $L=10$, $C=0.05$, $\theta=\pi/4$, with $R=0, 3, 6, 9$ in (\textit{a}-\textit{d}) respectively.
 The shaded inset solutions lie on the dashed branch of subharmonic steady solutions, which is unstable; all other solutions are harmonic, with period $L$.
  Solutions filled in black are stable, while white solutions are unstable.
We also indicate pitchfork bifurcations at $A=A_P$ ({\color{blue}$\blacklozenge$}), limit points at $A=A_{LP}$({\color{red}$\CIRCLE$}), Hopf bifurcations within the domain of length $L$ ($\blacksquare$), and states with $\min{q}=0$ ($\bigstar$).
\label{fig:Increasing_A_at_fixed_R}
 }
\end{figure}

As $A$ increases, the film profile deviates from the uniform state, and the nonlinearities in equations \eqref{eq:Benney} and \eqref{eq:WR} can lead to bifurcations between steady solutions.
Figure \ref{fig:Increasing_A_at_fixed_R} shows the behaviour of steady solutions to the weighted-residual equations as $A$ is varied for a selection of $R$ at fixed $L$, $\theta$ and $C$ subject to constrained mean layer height $1$.

 For each $R$, the only steady solution at $A=0$ is the uniform state with $h=1$. All steady solutions for $A>0$ are non-uniform, and the extent of this non-uniformity increases with $A$ when $A$ is moderate. However, for sufficiently large $A$, there are no steady solutions that describe continuous liquid films, due to a limit point in the bifurcation diagram.
 We can follow the steady solution branches around the limit point, and find that each branch eventually terminates when the minimum film height tends to zero, so that the layer dries up at some position.
 The film height profiles for these drying states are shown as insets in figure~\ref{fig:Increasing_A_at_fixed_R}.
 
 \begin{figure}
\includegraphics{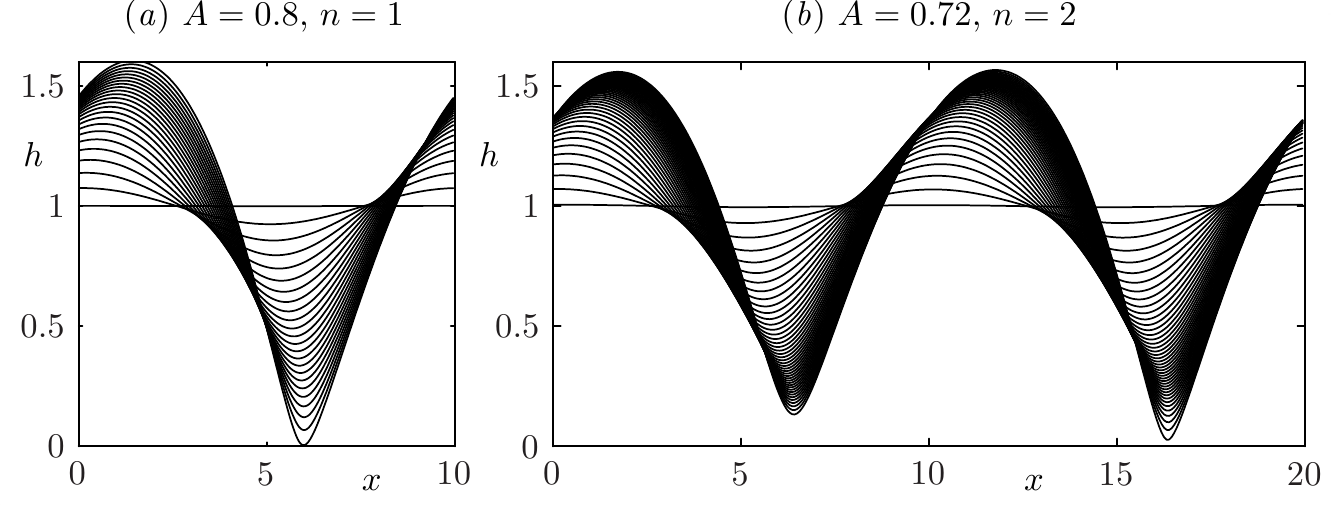}
 \caption{Time evolution from a near-flat initial state, for $R=0$, $L=10$, $C=0.05$, $F=A\cos(2\pi x/L)$, in (\textit{a}) a domain of length $L$ for $A>A_P$ so that there are no steady states, and (\textit{b}) a domain of length $2L$ with $A>A_P$ so that all steady states are unstable. In both cases, the film height vanishes in finite time at a single point in the domain.
 \label{fig:Subharmonic_crash}
 }
\end{figure}
 
 When $A$ is sufficiently large that there are no steady solutions, we explore the system behaviour by conducting time-dependent simulations in a periodic domain of length $L$, with a typical result shown in figure \ref{fig:Subharmonic_crash}(\textit{a}). We find that the film thins rapidly, and is able to dry in finite time due to the fixed speed of fluid removal. We have also conducted unsteady simulations starting from slightly-perturbed states on the solution branch with the lower minimum height. We find that these steady states are unstable, and the instability manifests as thinning and drying behaviour, rather than tending towards the linearly stable steady states with a larger minimum film thickness.

\begin{figure}
 \includegraphics{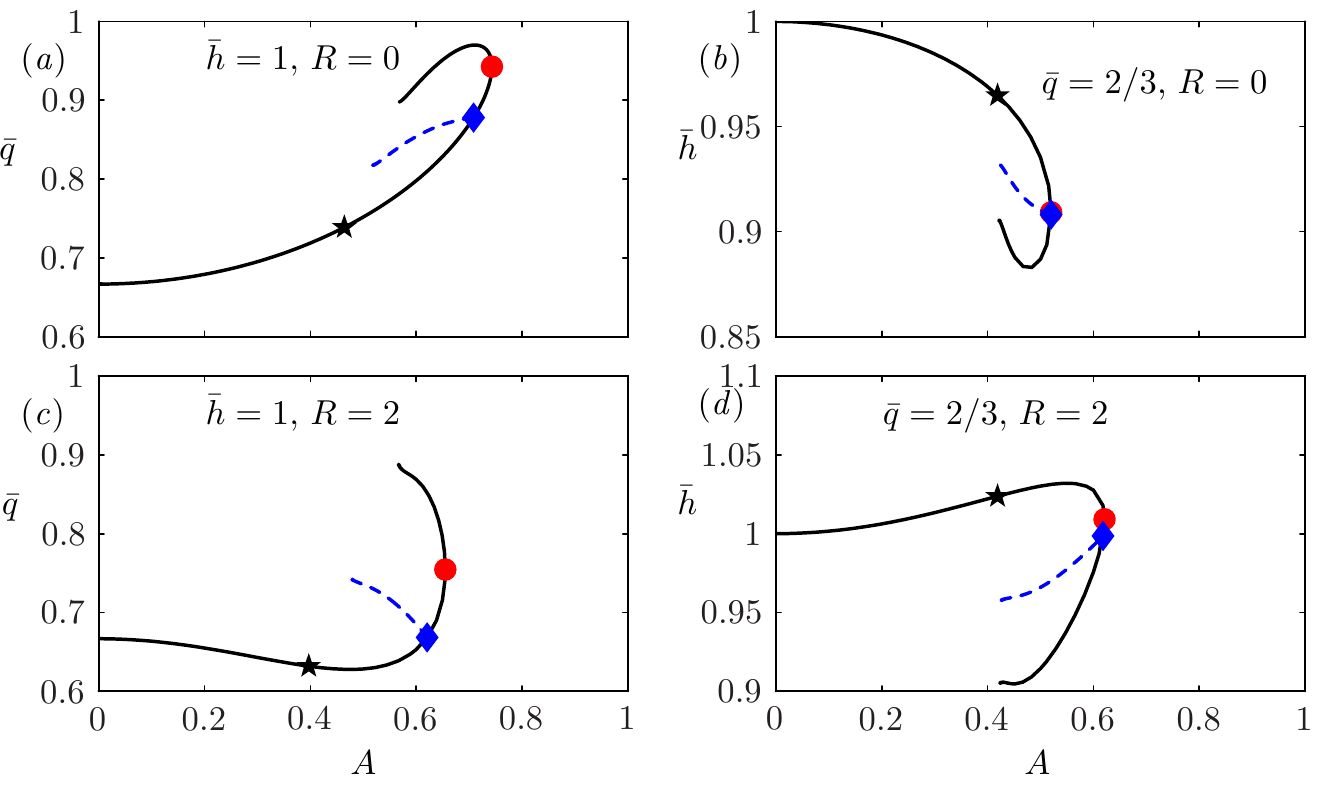}
 \caption{The bifurcation structure for steady solutions to the weighted residual equations as $A$ increases, subject to fixed mean height (\textit{a, c}) and fixed mean flux (\textit{b, d}). Here $L=10$, $C=0.05$ and $\theta=\pi/4$.
 The symbols mark the same bifurcations as in figure \ref{fig:Increasing_A_at_fixed_R}, but we do not show Hopf bifurcations here.
 Solution branches terminate at a point where the minimum layer height vanishes. 
 \label{fig:mean_flux} }
\end{figure}

The solutions shown in figure \ref{fig:Increasing_A_at_fixed_R} are all computed subject to a fixed mean layer height of 1, and are plotted according to the minimum value of $h$ in a single period. An alternative solution measure is the mean flux $\bar{q}$, averaged over one spatial period. 
Figure \ref{fig:mean_flux}(\textit{a}, \textit{c}) shows the bifurcation structure plotted in terms of the mean flux; we find that imposing blowing and suction at finite wavelength can either increase or decrease the mean flux. 
Figure \ref{fig:mean_flux}(\textit{b}, \textit{d}) shows bifurcation diagrams for two values of $R$ with fixed $\bar{q}=2/3$ and mean height free to vary (this condition is appropriate to experimental investigations performed under open conditions, corresponding to fixed mean down-slope flux). We find that, as was the case for the fixed $\bar{h}=1$ calculations, there is a limit point corresponding to a maximum value of $A$ with steady solutions.
However, time-dependent calculations for open conditions require non-periodic boundary conditions, which we have not pursued here.

\subsection{Subharmonic steady states}

The steady solutions discussed in \S~\ref{sec:Steady_increasing_A} are limited to cases where both the steady solution and the imposed blowing and suction are periodic with the same wavelength $L$. However, subharmonic steady solutions may also exist, for which the steady solution is spatially periodic with period $nL$, where $n$ is an integer equal to 2 or greater; we calculate subharmonic steady states by continuation in a domain of length $nL$.
For each case in figure \ref{fig:Increasing_A_at_fixed_R}, and also for the fixed $\bar{q}=2/3$ calculations shown in figure \ref{fig:mean_flux}, we have detected subharmonic solution branches with $n=2$.
In each of these cases, the subharmonic solutions emerge via a subcritical pitchfork bifurcation at $A=A_P$ in the steady solution structure, and the subharmonic steady states are all unstable. 
The unstable eigenmode of these states results in a drying process similar to that occurring for unstable harmonic steady states.

As the subharmonic steady states shown in figure \ref{fig:Increasing_A_at_fixed_R} are unstable, the only possible stable steady states are harmonic, with period $L$. However, for $A>A_P$, the period-$L$ steady states are also unstable to perturbations of wavelength $2L$; this is a consequence of the subcritical nature of the pitchfork bifurcation. The subharmonic instability eventually results in a film thinning and drying event, but at only one of the local minima within the domain of length $2L$. One such drying sequence is illustrated in figure \ref{fig:Subharmonic_crash}(\textit{b}).

\subsection{Streamfunction and flow reversal}
\label{sec:Streamfunction}

In addition to altering the interface height, the imposed suction
also fundamentally affects the arrangement of streamlines within the fluid film.
As the flow is incompressible, we can write the velocity in terms of a streamfunction $\psi(x,y)$
such that $(u,v) = (\psi_y, -\psi_x)$.
Under the weighted-residual formulation, we have
\begin{equation}
\label{eq:WR_velocity}
 u = \psi_y= \frac{3q}{h} \left( \frac{y}{h}-\frac{y^2}{2h^2}\right) +O(\delta) , \quad v = -\psi_x,
\end{equation}
with boundary condition $v = F(x)$ on $y=0$. Recalling that $F(x)=q'(x)$ for steady solutions, we can write the solution for $\psi$, up to the addition of an arbitrary constant, as
\begin{equation}
\label{eq:streamfunction}
 \psi = -q(x) + \frac{3q(x)}{h(x)}\left( \frac{y^2}{2h}-\frac{y^3}{6h^2}\right)+O(\delta).
\end{equation}
The same result applies to $O(\delta)$ in the Benney equations.

According to \eqref{eq:streamfunction}, there are no stagnation points in $0<y \leq h$ if $q>0$, and
points along the wall are stagnation points if and only if $F(x)=0$. In the absence of suction, every point on the wall is a stagnation point.
With non-zero blowing and suction, there are isolated stagnation points on the wall at the zeros of $F(x)$.

We now examine the steady flow near such an isolated stagnation point, located at $x=x_0$, $y=0$, with $F(x_0)=0$.
As $q'(x)=F(x)$, we can write
\begin{equation}
\label{Fq_streamfunction_expansions}
 q  \sim q(x_0) + F'(x_0)\frac{(x-x_0)^2}{2}+ O(x-x_0)^3.
\end{equation}
Substituting \eqref{Fq_streamfunction_expansions} into \eqref{eq:streamfunction} yields
\begin{equation}
 \psi \sim - q(x_0) - \frac{F'(x_0) x^2}{2} + \frac{3q(x_0)y^2}{2h(x_0)^2}+O((x-x_0)^2+(y-y_0)^2).
\end{equation}
If $F'(x_0) q(x_0)>0$, the stationary point at $x_0$ is a saddle point of $\psi$, and the streamlines are locally straight lines, such that
\begin{equation}
 \frac{y^2}{x^2} = \frac{F'(x_0) h(x_0)^2}{3q(x_0)}.
\end{equation}
The streamlines become increasingly vertical as $q(x_0)\rightarrow 0$.
In contrast, if $F'(x_0) q(x_0)<0$, the stationary point at $x_0$ is a local extremum of $\psi$, and the streamlines are locally elliptical, with
\begin{equation}
  -F'(x_0) x^2 + \frac{3q(x_0) y^2}{h(x_0)^2} = \text{const}.
\end{equation}
For any smooth, periodic, non-zero $F(x)$ with mean zero, there must be some stagnation points $x_0$ with $F'(x_0)>0$ and others with $F'(x_0)<0$.
 There are no steady solutions in figure \ref{fig:Increasing_A_at_fixed_R} with negative $h$,
 but the same is not true for $q$; for each $R$ with solutions shown in figure \ref{fig:Increasing_A_at_fixed_R}, there is an amplitude $A^*$ above which all steady solution have $q(x)<0$ over a finite interval in $x$.
As $q_x = F$, the minimum of $q$ occurs at a point $x_0$ with $F(x_0)=0$ and $F'(x_0)>0$.

\begin{figure}

\subfloat[Steady state for $A=0.2$]{\includegraphics[width=\textwidth]
{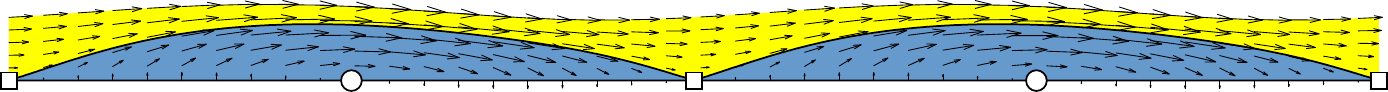}}

\subfloat[Steady state for $A=0.4$]{\includegraphics[width=\textwidth]
{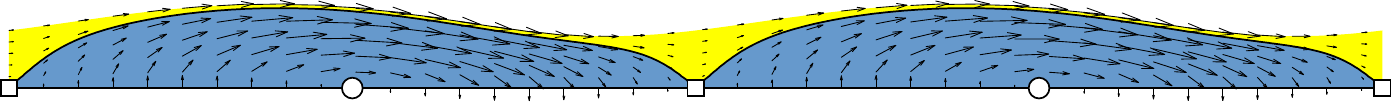}}

\subfloat[Steady state for $A=0.5$]{\includegraphics[width=\textwidth]
{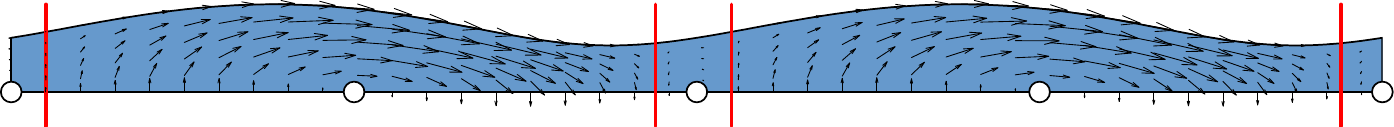}}

\subfloat[Steady state for $A=0.6$]{\includegraphics[width=\textwidth]
{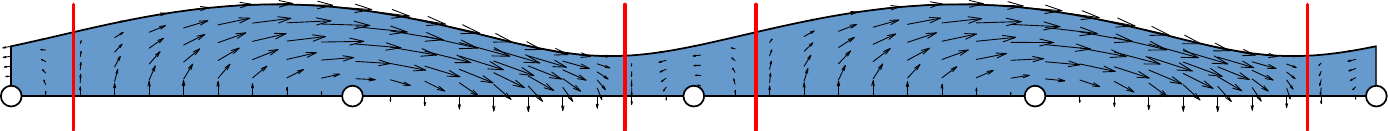}}

\subfloat[There are no steady states for $A=0.8$; this is a final snapshot before drying.]
{\includegraphics[width=\textwidth]{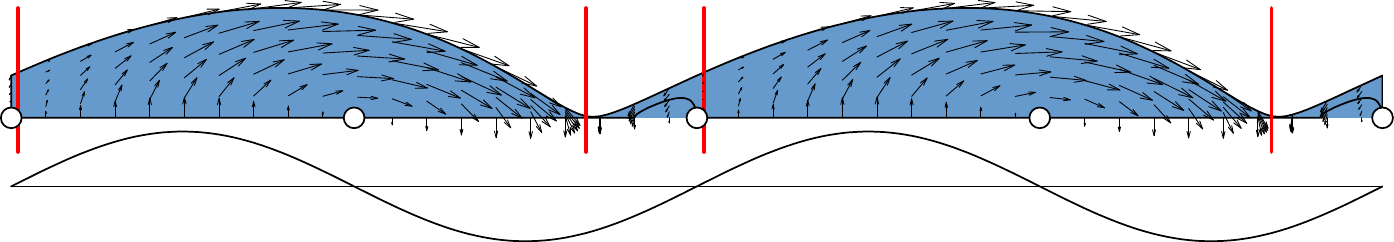}}

\caption{Solutions for the interface shape, velocity field and stagnation points for $\theta=\pi/4$, $C=0.05$, $L = 10$, $R=0$. Periodicity is enforced with $L=10$, but solutions are plotted over two periods, and are shown with aspect ratio 1. 
Stagnation points with $q(x)F'(x)<0$ and with $q(x)F'(x)>0$ are indicated by {\Large $\Circle$} and {\Large $\Square$} respectively.
In (\textit{a}-\textit{b}), $q>0$ everywhere, so a streamline emanating from the stagnation point with $q(x)F'(x)>0$ divides fluid into particles which never meet the wall (yellow), and particles which enters and leaves through the wall (blue). For $A>A^*=0.46$, all steady solutions have regions of negative $q$, corresponding to a region of upstream flow near the stagnation point (between the vertical lines), and all fluid particles must reach the wall.
\label{fig:Flow_reversal}
 }
\end{figure}

  There are no stagnation points inside the fluid domain, and so streamlines never cross. For small $A$, $q(x)>0$ for all $x$, and so stagnation points with $F'(x)>0$ are saddle points of $\psi$, while those with $F'(x)<0$ are extrema of $\psi$. Two examples of the flow field can be seen in figure \ref{fig:Flow_reversal}(\textit{a}, \textit{b}). Streamlines emanate into the fluid domain from each stagnation point with $F'(x)>0$, and these stagnation points are connected by a streamline which separates the fluid into a layer in which fluid particles propagate down the plane, and a layer in which fluid particles must enter and leave the flow domain via injection through the walls. 
  At $A=0$, all particles propagate. As $A$ increases, the propagating layer thins and the injection layer thickens.

At $A=A^*$, the stagnation point at $x=x_0$, $y=0$ switches from a saddle point to an extremum of $\psi$.
The corresponding change in the flow field is illustrated in the transition from figure \ref{fig:Flow_reversal}(\textit{b}) to \ref{fig:Flow_reversal}(\textit{c}). For $A>A^*$, $q(x)$ is negative for some $x$, and hence by \eqref{eq:WR_velocity} the horizontal velocity is directed up-slope for all $y$. As a result, no particles can propagate more than one period down the plane, and so the propagating layer vanishes for $A>A^*$.
 The width of the region with up-slope flow increases with $A$, until eventually there are no further steady solutions. Nonlinear time evolution calculations at large $A$ show that the film dries at a point just upstream of the negative-$q$ stagnation point. An instantaneous snapshot of the system at the moment of drying is shown in figure \ref{fig:Flow_reversal}(\textit{e}).
 
\section{Linear stability}
\label{sec:Linear_stability}

\subsection{Stability of uniform state}

In the absence of imposed suction, the only steady state with mean layer height unity is the Nusselt film solution, with $h=1$ and $q=2/3$. We linearise about this state, and seek eigenmodes proportional to $\exp(ikx+\lambda t)$. The Benney model yields the eigenvalue $\lambda$ directly, as 
\begin{equation}
\label{lambda_Benney}
 \lambda = -2ik +k^2 \left(\frac{8R}{15}-\frac{2}{3}\cot\theta\right)-\frac{k^4}{3C}.
\end{equation}
For the weighted residual model, the linear stability analysis yields a quadratic equation for $\lambda$:
\begin{equation}
\label{lambda_eq_WR}
\frac{2R}{5} \lambda^2 + \lambda \left( 1 -ik \frac{68R}{105}\right) + 2ik + k^2\left( \frac{2\cot\theta}{3} + \frac{k^2}{3C}-\frac{8R}{35}\right)=0.
\end{equation}
In both models,  the stability threshold for fixed $k$ is
\begin{equation}
\label{eq:RH_unforced}
 R < R_H \equiv \frac{5}{4} \cot\theta + \frac{5}{8C}k^2.
\end{equation}
The linear stability threshold \eqref{eq:RH_unforced} is also valid for perturbations with long wavelengths in full solutions to the Navier--Stokes equations \citep{Benjamin}.
At $R=R_H$, both \eqref{lambda_Benney} and \eqref{lambda_eq_WR} have a root with negative real part for $R<R_H$, the value $\lambda=\lambda_0=-2ik$ at $R=R_H$, and positive real part for $R>R_H$.
We note that \eqref{lambda_eq_WR} also has a second root for $\lambda$, but this root always has negative real part when $R>0$.

\subsection{Perturbations of arbitrary wavelength about a periodic base state}
\label{sec:Infinite_stability_methods}

When the base state is spatially uniform, the eigenmodes can be written as $\Re(\exp(ikx+\lambda t))$, and by calculating $\lambda$ for all real $k$, we can determine linear stability to perturbations of all wavelengths. However, when non-zero suction and blowing is imposed, the base state is not uniform, and so the eigenmodes are no longer simple exponential functions. Fortunately, Floquet-Bloch theory allows us to compactly describe eigenmodes of arbitrary wavelength.

We suppose that we are given a steady solution $h=H(x)$, $q=Q(x)$ to the forced equations, and consider the evolution of arbitrary small perturbations, so that
\begin{equation}
 h(x,t) = H(x) + \epsilon\Re\lbrace\hat{h}(x,t)\rbrace+O(\epsilon^2), 
 \quad q(x,t) = Q(x) + \epsilon\Re\lbrace\hat{q}(x,t)\rbrace+O(\epsilon^2), 
\end{equation}
with $\epsilon \ll 1$.
The mass conservation equation \eqref{eq:mass} becomes
\begin{equation}
\label{stability_h}
 \hat{h}_t + \hat{q}_x=0.
\end{equation}
At $O(\epsilon)$, the Benney flux equation \eqref{eq:Benney} yields
\begin{equation}
\label{stability_q}
 \hat{q} = b_0(x) \hat{h} + b_1(x) \hat{h}_x + b_2(x) \hat{h}_{xxx}
\end{equation}
where
\begin{equation}
\label{b0_stability}
 b_0 = H^2 \left(2-2 H_x\cot\theta + \frac{H_{xxx}}{C}\right) + \frac{16 R}{5}H^5 H_x - \frac{8R}{3} H^4 F,
\end{equation}
\begin{equation}
\label{b12_stability}
 b_1 = -\frac{2}{3}H^3 \cot\theta + \frac{8}{15}R H^6, \quad b_2 = \frac{H^3}{3C}.
\end{equation}
The weighted residual equivalent of \eqref{stability_q} additionally involves a term proportional to $\hat{q}_t$.
Given a base solution $H(x)$, $Q(x)$ and $F(x)$, the coefficients $b_0$, $b_1$ and $b_2$ are all known periodic functions of $x$, with the same period as the base solution.

We now invoke the Floquet-Bloch form, observing that as \eqref{stability_h} and \eqref{stability_q} are linear equations with coefficients that are periodic in $x$ with period $L$,
the
eigenfunctions can be written as 
\begin{equation}
\label{Floquet_eigenfunctions}
 \hat{h}(x,t) = e^{\lambda t + i kx}\hat{h}_k(x), \quad \hat{q}(x,t)  = e^{\lambda t + ikx} \hat{q}_k(x),
\end{equation}
where the Floquet wavenumber $k$ is real, and $\hat{h}_k$ and $\hat{q}_k$ are periodic functions of $x$ with period $L$.
 Setting $k=0$ recovers perturbations of wavelength $L$, while very small but non-zero $k$ corresponds to very long wave perturbations. 
The solution is stable to perturbations with period $L$ if $\Re(\lambda)<0$ for all eigenfunctions when $k=0$.
The base solution is linearly stable to perturbations of all wavelengths if $\Re(\lambda)<0$ for all eigenfunctions for each real $k \in [0, \pi/L]$.

\subsection{Effect of small-amplitude blowing and suction on stability}

We now analyse the effect of small amplitude blowing and suction in the form $F=A\cos{mx}$ on the stability of eigenmodes with underlying Floquet wavenumber $k$, which will allow us to determine how such forcing affects the critical Reynolds number. We do so by expanding the eigenvalue $\lambda$ for $R$ close to $R_H$, and for small $A$:
\begin{equation}
\label{lambda_expansion}
 \lambda = \lambda_0 + (R-R_H)\left.\frac{\partial \lambda}{\partial R}\right.+ Z A^2 +O( (R-R_H)^2) + O((R-R_H)A^2) + O(A^4).
\end{equation}
We are concerned with eigenvalue equal to $\lambda_0 = -2ik$ at $R=R_H$, which is a root of both the Benney and weighted residual characteristic equations.
 We can evaluate the term $\partial \lambda/\partial R$ which appears in \eqref{lambda_expansion} by differentiating \eqref{lambda_Benney} or \eqref{lambda_eq_WR} as appropriate.
The eigenvalue is even in $A$ because the transformation $A\rightarrow -A$ can be recovered by translation in $x$ by a distance $L/2$, and the original eigenmode $\exp(ikx+\lambda t)$ has no preferred position; thus the leading order contribution for small $A$ is $O(A^2)$.

The effect of the imposed suction is encapsulated in the coefficient $Z$, which depends on the suction wavenumber $m$, the perturbation wavenumber $k$, and $C$ and $\theta$. 
In order to calculate $W$ and $Z$, we need to calculate both the base solution and the eigenfunction to $O(A^2)$.
The calculation of the base solution to $O(A^2)$ was discussed earlier in the context of steady states, and the required expansion is given by \eqref{Q_steady_expansion} and \eqref{H_steady_expansion}.

For small-amplitude steady solutions, the suction function $F$ and the base solution $H$, $Q$ are all periodic with wavenumber $m$, and so the coefficients in the linearised equations \eqref{stability_h} and \eqref{stability_q} are also periodic with wavenumber $m$.
As a result, all eigenfunctions of \eqref{stability_h} and \eqref{stability_q} can be written in the Floquet form given by \eqref{Floquet_eigenfunctions}.
The unknown functions $\hat{h}_k$ and $\hat{q}_k$ are periodic with wavenumber $m$ and are constant when $A=0$, and can be expanded for small $A$ as
\begin{equation}
\label{h_star_expansion}
 \begin{array}{ll}
 \hat{h}_k(x) =  1  &+ \quad  A (C_{1} e^{imx} + C_2+C_{3} e^{-imx})  \\[8pt]
    &  + \quad  A^2 (D_1 e^{2imx} + D_2 e^{imx} + D_3 + D_4 e^{-imx} +D_5 e^{-2imx}) 
 \end{array}
\end{equation}
and
\begin{equation}
\label{q_star_expansion}
 \begin{array}{ll}
 \hat{q}_k(x) = \dfrac{i\lambda_0}{k}  &+ \quad A (E_{1} e^{imx} + E_2 +E_{3} e^{-imx})  \\[8pt]
    & + \quad  A^2 (G_1 e^{2imx} + G_2 e^{imx} + G_3 + G_4 e^{-imx} +G_5 e^{-2imx}),
 \end{array}
\end{equation}
where the complex constants $C_1$, $C_2$, $C_3$, $D_1$, $D_2$, $D_3$, $D_4$, $D_5$, $E_1$, $E_2$, $E_3$, $G_1$, $G_2$, $G_3$, $G_4$ and $G_5$ are to be found along with $Z$.
We must also choose a normalisation condition on the amplitude and phase of the eigenvector, and for this we use the condition
\begin{equation}
\label{evec_normalisation}
 \hat{h}_k(0)=1.
\end{equation}
To determine the unknown constants, we set $R=R_H$, solve for the steady state given by \eqref{Q_steady_expansion} and \eqref{H_steady_expansion}, and then substitute the eigenvalue expansion from \eqref{lambda_expansion} and the eigenfunction from \eqref{Floquet_eigenfunctions}, \eqref{h_star_expansion}, \eqref{q_star_expansion} into the two linearised equations \eqref{stability_h} and \eqref{stability_q}, and also the normalisation condition \eqref{evec_normalisation}. We solve the linearised equations first at $O(A)$ and then at $O(A^2)$, at each order obtaining linear systems for the unknown coefficients including $Z$.
We perform this calculation in Maple, and obtain a lengthy expression for $Z$ as a function of $m$, $n$, $C$ and $\theta$; the full result depends on whether we have used the weighted-residual or Benney equations.
Once $Z$ is known, we can obtain the neutral stability curve for fixed $k$, defined by the condition $\Re\lbrace \lambda \rbrace=0$, so that 
\begin{equation}
\label{W_definition}
 R \sim R_H  + A^2 W, \quad W \equiv -\Re\lbrace Z \rbrace
 \left(\Re \left\lbrace \dfrac{\partial \lambda}{\partial R}\right\rbrace\right)^{-1}.
\end{equation}

\begin{figure}
 \includegraphics{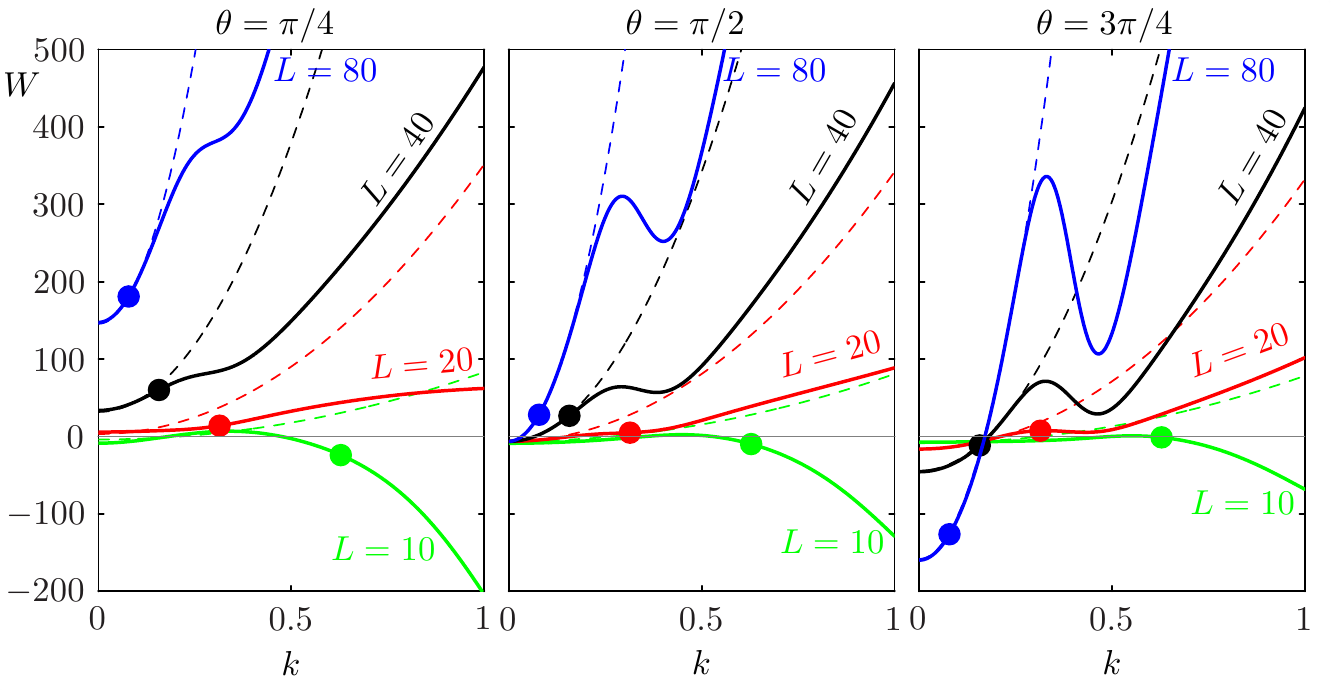}
 \caption{Effect of small $A$ blowing and suction with wavelength $L$ on the critical Reynolds number for perturbations of underlying wavenumber $k$ as measured through the weighted-residual results for $W$, with $C=0.05$. The filled circles on each curve mark the point where $k=2\pi/L$. 
The dashed lines show the long-wave asymptotic estimate \eqref{W_expansion_Benney}.
 \label{fig:Spectrum_comparison}
 }
\end{figure}

Figure \ref{fig:Spectrum_comparison} shows the effect of small $A$ on the stability boundary, as quantified by the weighted-residual results for $W$. 
For $\theta=\pi/4$, we find that forcing via blowing and suction at wavelength $L\equiv 2\pi/m=10$ destabilises short wavelength perturbations, but has little effect on long wavelength perturbations. 
However if the forcing wavelength $L\geq 20$, $W$ is positive for all $k$, and so forcing has a stabilising effect on all perturbations. Furthermore, for long wavelength forcing, the magnitude of $W$ increases with $L$, and the minimum value of $W$ occurs at $k=0$; this value is positive and increases rapidly with $L$.
For $\theta=\pi/2$, forcing with wavelength $L$ has a destabilising effect on short waves for $L=10$, and a stabilising effect for $L=20,$ 40, 80. However, $W$ tends to approximately $-6$ as $k\rightarrow 0$, and so the imposed forcing in fact slightly destabilises long wave perturbations.
For $\theta=3\pi/4$, the behaviour for small perturbation wavenumber is reversed from the $\theta =\pi/4$ case; long wave forcing is always destabilising for $k=0$, and becomes increasingly destabilising as $L$ is increased.

\begin{figure}
 \includegraphics{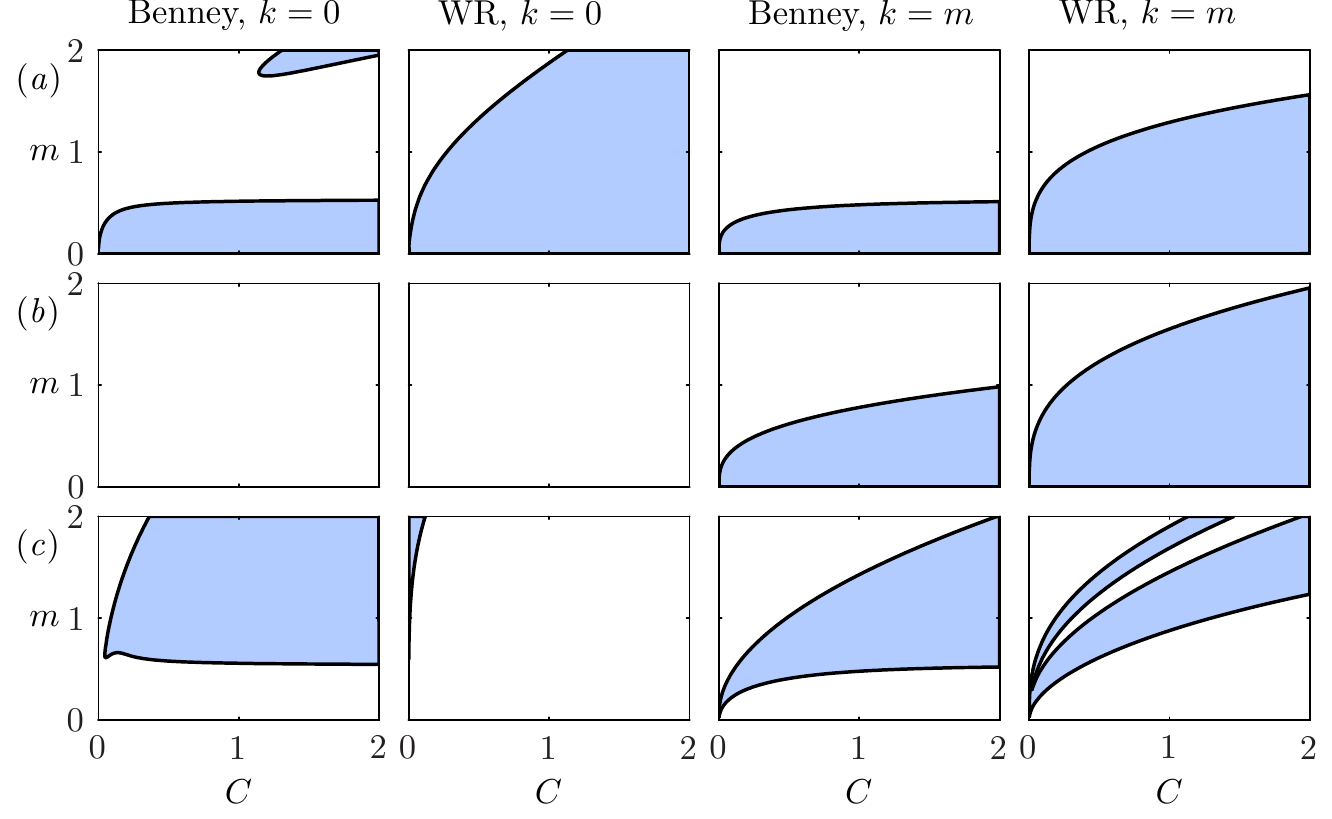}
 \caption{The effect of small amplitude blowing and suction with wavenumber $m$ on perturbations with $k=0$ and with $k=m$ in the Benney and weighted-residual (WR) models, as described by the quantity $W$ defined in \eqref{W_definition}; $W>0$ in the shaded regions, so small-amplitude blowing and suction increases the critical Reynolds number, and has a stabilising effect on the flow.
 Results for $\theta=\pi/4, \pi/2, 3\pi/4$ are shown in rows (\textit{a}-\textit{c}) respectively.
 \label{fig:sigma_2_comparison} }
\end{figure}

In figure \ref{fig:sigma_2_comparison}, we survey the sign of $W$ for $k=0$ and $k=m$, with $m$ and $C$ varying, for three values of $\theta$, for both the Benney and weighted-residual models.
 The two models concur that for each inclination angle $\theta$, there are some regions of $(C,m)$ parameter space where the imposition of blowing and suction stabilises the uniform state, and other regions where such forcing destabilises the flow.
 For $\theta=\pi/4$, both models predict that long wave forcing, with $m\rightarrow 0$, is stabilising to perturbations with Floquet wavenumber $k=0$ and with wavenumber $k=m$. For $\theta=\pi/2$, forcing is destabilising in the limit $k\rightarrow 0$ for all $C$ and $m$, but has a stabilising effect on perturbations with $k=m$ if $m$ is not too large. For $\theta=3\pi/4$, the long wave forcing is destabilising to perturbations both with $k=0$ and with $k=m$.

 The critical Reynolds number increases quadratically with perturbation wavenumber $k$ in the absence of blowing and suction, and so for very small $A$, the lowest critical Reynolds number must also be obtained at small $k$.
We now consider $W$ in the limit of long-wavelength blowing and suction and long perturbation wavelength, setting
$m=\delta M$, $k=\delta K$ and $C = \delta^2 \widehat{C}$, and expanding the full expression for $Z$ with $\delta \ll 1$.
Both the Benney and weighted-residual results yield
\begin{equation}
\label{Z_expansion}
  Z \sim \frac{3 i K}{4\delta M^2} + \left[-\frac{K^2 \cot\theta}{2M^2} + \frac{K^2}{6\widehat{C}} - \frac{11 K^4}{12 \widehat{C} M^2}\right]+ O(\delta^2), \quad
 \Re \left\lbrace \frac{\partial \lambda}{\partial R}\right\rbrace \sim \frac{8K^2\delta^2}{15} + O(\delta^4).
\end{equation}
Substituting \eqref{Z_expansion} into \eqref{W_definition} yields the leading order long-wave expansion
\begin{equation}
\label{W_expansion_Benney}
 W  = -\frac{15}{8\delta^2}\left[-\frac{\cot\theta}{2M^2} + \frac{1}{6\widehat{C}} - \frac{11 K^2}{12 \widehat{C} M^2}\right] +O(1)
 =-\frac{15}{8}\left[-\frac{\cot\theta}{2m^2} + \frac{1}{6 C} - \frac{11 k^2}{12 C m^2}\right] +O(1).
\end{equation}
The prediction \eqref{W_expansion_Benney}
is plotted in figure \ref{fig:Spectrum_comparison} for comparison to the non-long-wave results obtained by direct evaluation of the Maple expressions for $Z$ and $\partial \lambda/\partial R$.

For small-amplitude blowing and suction at fixed wavenumber $m$, the critical Reynolds number for instability to perturbations of all real wavenumbers $k$ is
\begin{equation}
\label{Rstar_definition}
 R^*(m) = \min_{k\in \mathbb{R}} \left[ \frac{5}{4}\cot\theta + \frac{5k^2}{8C}+A^2 W(m,k,C, \theta) \right].
\end{equation}
Using the long wave expansion for $W$ given by \eqref{W_expansion_Benney}, we obtain
\begin{equation}
\label{eq:R_star_middle}
 R^*(m) =  \frac{5}{4}\cot\theta +\frac{15A^2}{16m^2}\cot\theta 
 -\frac{5A^2}{16C} +\min_{k\in \mathbb{R}} \left[ k^2 \left(\frac{5}{8C}
 + \frac{55A^2 }{32m^2 C}\right)
 \right].
\end{equation}
The term in square brackets is positive, and so the minimum of the expression occurs at $k=0$. As expected, this implies that long-wave perturbations are the first to become unstable as $R$ is increased, both for $A=0$ and also for small $A$ in the long wave limit.

Proceeding with the long wave approximation, we evaluate the minimum of \eqref{eq:R_star_middle} as
\begin{equation}
\label{eq:R_star_final}
 R^* \sim \frac{5}{4}\cot\theta + \frac{15}{16}A^2 \left(\frac{\cot\theta}{m^2}-\frac{1}{3C}\right),
\end{equation}
while if perturbations 
restricted are restricted to those with wavenumber $k=m$, we find
\begin{equation}
\label{eq:R_star_restricted}
 R_{m=k} \sim  \frac{5}{4}\cot\theta + \frac{5m^2}{8C}
 + \frac{15}{16}A^2 \left( \frac{\cot\theta}{m^2} -\frac{3}{2C}\right).
\end{equation}
The stability boundaries for perturbations with all $k$ and with $k=m$ are plotted in figure \ref{fig:RA_maps} for nonlinear solutions to the weighted-residual equations, small $A$ predictions obtained using the full version of $W$, and the long-wave, small $A$ predictions \eqref{eq:R_star_final} and \eqref{eq:R_star_restricted}. We see that the first two boundaries are in good agreement with each other when $A$ is sufficiently small, for both $L=40$ and $L=10$.
The long-wave small-amplitude predictions are in good agreement with the other two versions of boundaries when $L=40$, but are poor agreement when $L=10$. 

In the case $\theta=\pi/4$, $R^*$ is positive in the absence of suction, and long-wave blowing and suction can either decrease or increase $R^*$, depending on the values of $\theta$, $m$ and $C$, and here it is reasonable to consider an optimisation strategy.
To maximise the stabilising effect at fixed $A$ for $\theta<\pi/2$, we should choose $m$ as small as possible, and in fact \eqref{eq:R_star_final} predicts that $R^*\rightarrow +\infty$ as $m\rightarrow 0$ with $A$ fixed. 
% Imposing blowing and suction at short wavelengths, with $m^2>3C\cot\theta$, reduces the critical Reynolds number for stability to perturbations of all wavelengths.
Regarding constraints on the magnitude of film height variations, we note that 
these are of order $A/m$ for $A\ll 1$ (see \S~\ref{sec:Small_A}) , and so we can approximately constrain height variations by keeping $\alpha=A/m$ fixed. We find that the largest $R^*$ is again obtained as $m\rightarrow 0$, but in contrast to the fixed $A$ case, $R^*$ is bounded as $m\rightarrow 0$.

The governing equations are valid for $0<\theta<\pi$, with $\theta=\pi/2$ corresponding to flow down a vertical wall, and $\pi/2<\theta<\pi$ corresponding to flow along the underside of an inclined plane. In the absence of blowing and suction, $R^*=0$ for $\theta=\pi/2$, and $R^*<0$ for $\pi/2<\theta <\pi$. 
Negative Reynolds numbers are not physically relevant, but consideration of \eqref{eq:R_star_final} may still give some indication of the effect of blowing and suction when $\theta>\pi/2$.
% for inclinations angles greater than $\pi/2$.
According to \eqref{eq:R_star_final}, when $\theta \geq \pi/2$, imposing blowing and suction at long wavelengths always reduces $R^*$, and so is destabilising. 
Our other results for $\theta=3\pi/4$, shown in figures \ref{fig:Spectrum_comparison} and \ref{fig:sigma_2_comparison}, also suggest that suction destabilises long wave perturbations.

 \subsection{Finite $A$ stability results}

\begin{figure}

\subfloat[$L=10$, with $C=0.05$, $\theta = \pi/4$]{ \includegraphics{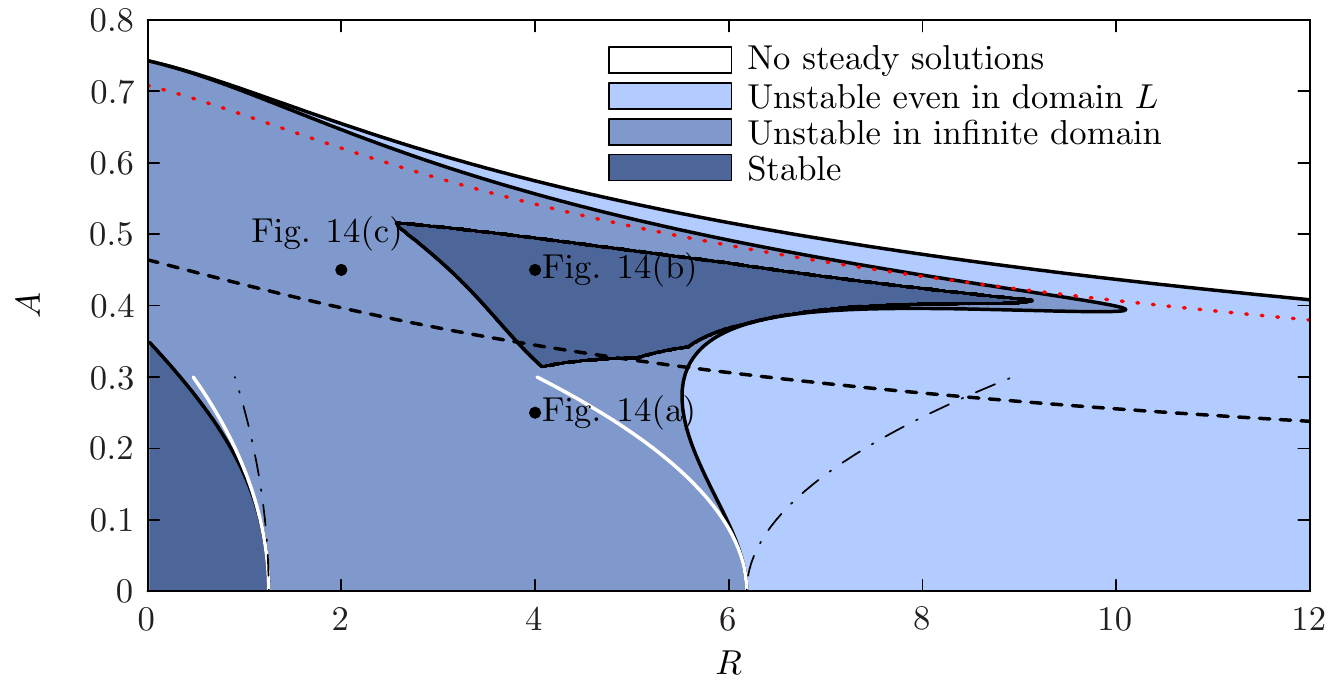}}

\subfloat[$L=40$, with $C=0.05$, $\theta = \pi/4$]{ \includegraphics{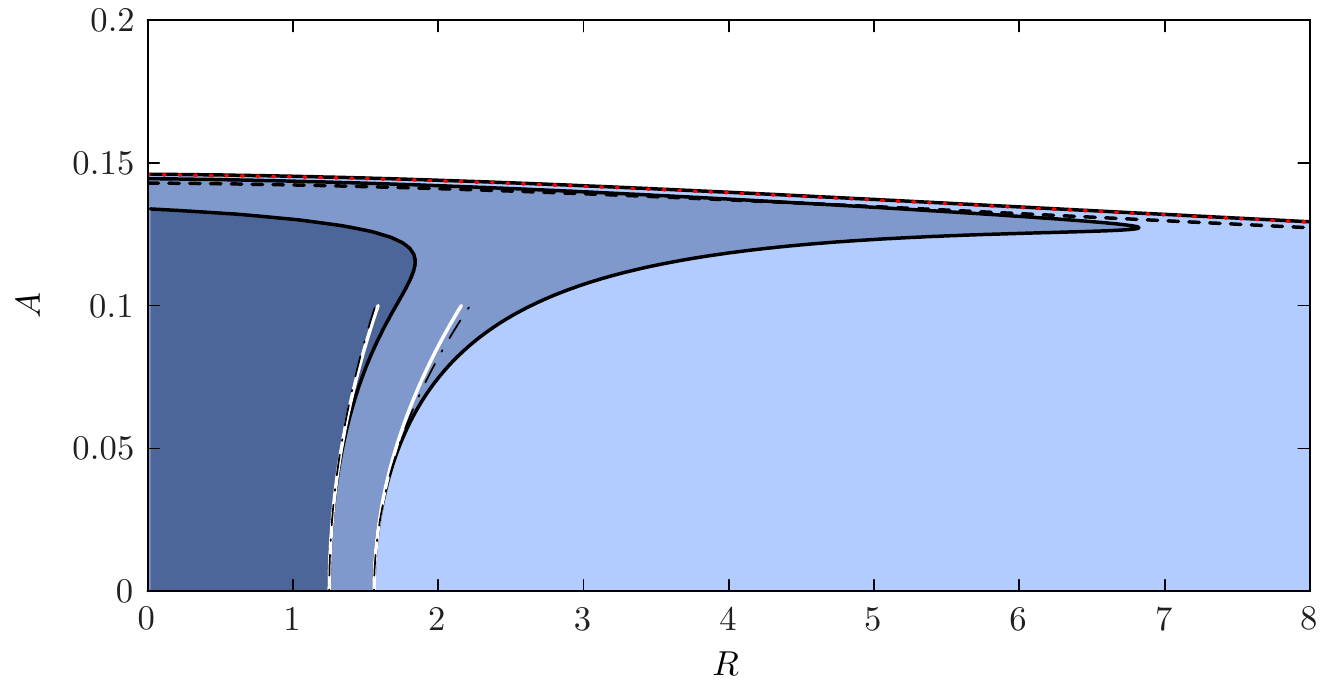}}

\caption{Stability properties for steady solutions as $R$ and $A$ vary at fixed $\theta=\pi/4$ and $C=0.05$, for the weighted-residual model. Steady solutions are divided into three linear stability categories as indicated by the legend, 
with finite-$A$ stability boundaries shown by solid black lines.
Two small-$A$ estimates for these boundaries are shown: the full result correct to $O(A^2)$ (dash-dotted line), and the long-wave, small $A$ estimates
\eqref{eq:R_star_final} and \eqref{eq:R_star_restricted} (solid white line). 
We also indicate solutions with $\min(q)=0$ (dashed line), and the pitchfork bifurcation $A_P$ (dotted line).
\label{fig:RA_maps}
}
\end{figure}

Figure \ref{fig:RA_maps} shows stability regions as $R$ and $A$ vary, at fixed values of $C$ and $\theta$, for imposed suction with wavelength $L=10$ and $L=40$. 
For $L=10$, introducing finite amplitude blowing and suction decreases the critical $R$ for linear stability to perturbations of wavelength $L$ and to perturbations of arbitrary wavelength. In contrast, when the imposed suction has wavelength $L=40$, increasing $A$ from zero initially increases both of these critical Reynolds numbers, and so has a stabilising effect on the base flow.

The boundary of the region stable to perturbations of fixed wavelength $L$ is defined by a Hopf bifurcation. When $A=0$, this is the Hopf bifurcation at $R=R_H$, for which the eigenmode can be written as $\hat{h}=e^{ikx}$ with $k=2\pi/L$. We can use \textsc{Auto-07p} to track the Hopf bifurcation as $A$ varies.  As $A$ is increased from zero, the eigenmode is modulated and is no longer a single Fourier mode, but remains periodic with period $L$.

Calculation of the stability boundary in the presence of patterned blowing and suction for perturbations of all wavelengths is not as simple as the case of tracking a single bifurcation, as we do not know \textit{a priori} which wavelengths are most unstable.
However, in the absence of blowing and suction, we know that long waves, with $0<k^2\ll 1$, are the first to become unstable as $R$ is increased. For $R>R_H$, there is a wavenumber cutoff, so that perturbations with $k^2<k^2_c$ are unstable, and there is a finite $k^2$ with maximum growth rate $\Re(\lambda)$. However, it is possible that for larger $A$, imposing blowing and suction can destabilise at some wavelengths while stabilising at others, and so the system may first become unstable at a non-zero wavenumber.
For finite $A$ we conduct a brute force computation of stability properties over a large number of perturbation wavelengths to determine stability with respect to perturbations of all wavelengths.

For the cases shown in figure \ref{fig:RA_maps}, we find that at small $A$, the boundary for stability to perturbations of arbitrary wavelengths agrees well with the asymptotic results derived in the previous subsection. For both $L=10$ and $L=40$, this boundary connects smoothly to a finite suction amplitude $A$ at $R=0$. There is a turning point with respect to $R$ in the $L=40$ results, and so the largest $R$ at which the system is stable occurs for a non-zero $A$. However, for $L=10$, there is no turning point, but there is a second `island' region where the steady state is stable to perturbations of all wavelengths. This region requires moderately large $A$ and $R$. Results from time dependent simulations in and around the island are shown in figure \ref{fig:stable_region}. As the island region occurs only for $L=10$, it is possible that it is simply an artefact of applying long-wave models at relatively short wavelengths.

 \subsection{Optimal wavelength for blowing and suction to obtain a stable steady state}
 \label{sec:Optimal_search}
 
 \begin{figure}
 \includegraphics{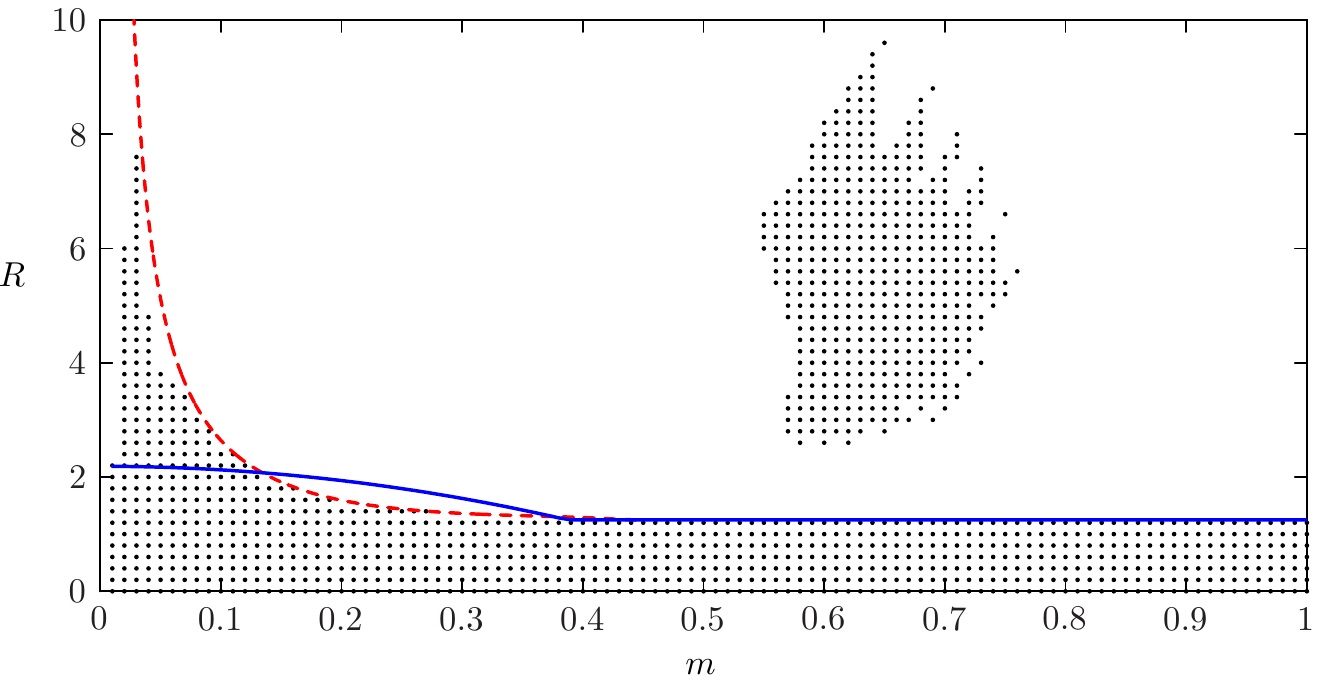}
 \caption{Finite $A$ stability results across a range of blowing and suction wavenumber $m$, with $F= A\cos{mx}$, $C=0.05$ and $\theta=\pi/4$. Each black
indicates that for the given $R$ and $m$, there is some $A$ for which there is a steady state stable to linear perturbations of all wavelengths. 
 The dashed line is obtained by explicit tracking of the maximum stable $R$ for fixed Floquet number $k=0.01$; this curve follows the stability boundary relatively well until it diverges at $m \approx 0.05$. Along this tracked curve, $\alpha=A/m$ is between $0.5$ and $1$. The blue line shows the small-$A$, long-wave stability prediction \eqref{eq:R_star_final} with $\alpha=1$, but this prediction underestimates the range of stable $R$ for small $m$.
 \label{fig:RM_search_WR}
 }
\end{figure}

 Determination of the largest $R$ that can be stabilised in an infinite domain by imposing a suitable suction profile is a complicated question, requiring finite amplitude results for the steady solutions and for their linear stability operator. 
 We can address this task numerically, by choosing a suction wavenumber $m$, increasing $R$ slightly from $R_0=1.25\cot\theta$, and testing the stability of steady solutions beginning from $A=0$ until $A$ is too large for steady states. 
Figure \ref{fig:RM_search_WR} shows numerical results from this search for the case $\theta=\pi/4$, $C=0.05$, using the weighted-residual model. We find two ranges of $m$ in which imposing suction can increase the critical Reynolds number.

Firstly, at large wavelengths with $L\approx 200$, imposing blowing and suction allows the critical Reynolds number for stability to all perturbations to increase from the unforced value of 1.25 to around 8.
However, figure \ref{fig:RM_search_WR} seems to indicate a downturn in the maximum stable $R$ at very small $m$. 
From the small $A$, long-wave result \eqref{eq:R_star_final}, we expect forcing via blowing and suction at small $m$ to have some stabilising effect, but this result does not predict the magnitude of the stabilising effect on its own, as we do not know $A$.
Furthermore, estimates of the maximum stable $R$ typically require large $A$ calculations, and so predictions based on the long-wave, small-amplitude expression \eqref{eq:R_star_final} are not particularly useful in this case.

The second region in which steady solutions are stable to perturbations of all wavelengths in figure \ref{fig:RM_search_WR} appears when the wavenumber $m$ for blowing and suction is relatively large, and only at moderately large $R$. This region corresponds to the island region in the $L=10$ results in figure \ref{fig:RA_maps}, but there is no such stable island in the results for $L=40$.

\section{Travelling waves}
\label{sec:Travelling_waves}

\subsection{Travelling waves in the absence of suction}

In the absence of suction, the system can support travelling waves, propagating at a constant speed $U$ without changing form. We can write the solution as 
\begin{equation}
 h(x,t) = H(\zeta), \quad q(x,t) = Q(\zeta), \quad \zeta = x - Ut,
\end{equation}
where $U$ is the unknown constant wave speed. $H$ and $Q$ must satisfy
\begin{equation}
 -U H' + Q'=0
\end{equation}
and
\begin{equation}
 -U Q' + Q =   \frac{H^3}{3}\left(2-2 H' \cot\theta + \frac{H'''}{C}\right) 
+
R
\left(\frac{18}{35} Q^2 H' - \frac{34}{35}Q Q' H\right),
\end{equation}
where a prime indicates a derivative with respect to $\zeta$.
If $H(\zeta)$ is periodic with period $L$, then the travelling wave solution is spatially periodic with period $L$, and temporally periodic with period $T = L/U$. 
We can compute large amplitude travelling waves numerically; figure~\ref{fig:Travelling_wave} shows one such travelling wave solution for $\theta=\pi/4$, $C=0.05$, $R=3$.

Individual travelling waves may be stable or unstable, and branches of travelling waves can undergo a range of bifurcations. However, small amplitude travelling waves connect to the uniform film state via a Hopf bifurcation at $R= R_H$.
This Hopf bifurcation can be supercritical or subcritical, with stable, small-amplitude travelling waves observed near the bifurcation only in the supercritical case.
We can determine the criticality of the Hopf-bifurcation by solving for small-amplitude limit cycles near to the critical Reynolds number via the following expansion :
\begin{equation}
\zeta = x - Ut,
\quad
 H(\zeta) = 1+ \epsilon \cos{k\zeta} + \epsilon^2 H_2(\zeta) + O(\epsilon^3), 
\end{equation}
\begin{equation}
 H_2(\zeta) = r_1 \cos{2k\zeta} + r_2 \sin{2k\zeta}, \quad q(x,t) = U H(\zeta) + S,
\end{equation}
\begin{equation}
 U = U_0 + \epsilon^2 U_2 + O(\epsilon^4), \quad S = S_0 + \epsilon^2 S_2 + O(\epsilon^4), \quad R = R_H + \epsilon^2 \bar{R} + O(\epsilon^4).
\end{equation}
We expand the equations for $\epsilon \ll 1$, and must solve the equations at up to $O(\epsilon^3)$ in order to determine $\bar{R}$.
We find that small amplitude travelling waves always travel downstream, with speed $U=U_0=2$, which is twice the velocity of particles on the surface.
The bifurcation is supercritical if $\bar{R}>0$ and subcritical if $\bar{R}<0$. The Benney equations yield
\begin{equation}
\label{eq:Rbar_Benney}
 \bar{R} = \frac{120C^2 - 60C^2 k^2 \cot^2 \theta  -120 Ck^4\cot\theta - 45k^6}{64 Ck^4}, 
\end{equation}
which may be positive or negative.
However, for the weighted-residual equations, we find
\begin{equation}
 \label{eq:Rbar_WR} 
 \bar{R} = \frac{ 4410 C^2 + 6670 k^2C^2 \cot^2\theta + 12235 k^4 C \cot\theta + 4450 k^6}{2352 C k^4}.
\end{equation}
When $\theta<\pi/2$, \eqref{eq:Rbar_WR} yields $\bar{R}>0$ and so the Hopf bifurcation in the weighted residual model is supercritical.
In the long-wave limit, with $k=\delta K$ and $C = \delta^2 \widehat{C}$, both \eqref{eq:Rbar_Benney} and \eqref{eq:Rbar_WR} yield 
\begin{equation}
\bar{R}  \sim \frac{15 \widehat{C} }{8\delta^2  K^4} +O(\delta^2) = \frac{15 C}{8 k^4} + O(\delta^2),
\end{equation}
which is positive.

The quantity $\bar{R}$ determines the criticality of the Hopf bifurcation, and so also governs the stability of small-amplitude travelling waves in the neighbourhood of the bifurcation. However, the branch of travelling waves may undergo further bifurcations \citep{Scheid_et_al}, and so the value of $\bar{R}$ does not necessarily prescribe the stability of finite-amplitude travelling waves.

\subsection{Influence of heterogeneous blowing and suction on travelling waves}
\label{sec:Perturbed_travelling_waves}

Introducing spatially-periodic suction means that the system is no longer translationally invariant, and so we cannot obtain true travelling waves for non-zero $A$. 
For the final part of our analysis, we calculate the effect of small amplitude suction on travelling waves, and consider in particular how travelling waves can transition to steady, stable, non-uniform states as the amplitude of the imposed blowing and suction is increased.

We now perturb the travelling wave by introducing a small-amplitude suction $F = A f(x)$ with $f(x)$ periodic with period $L_F$, and $A$ small.
We expect to find
\begin{equation}
\label{Travelling_perturbation_expansion}
 h(x,t) = H(\zeta) + A\hat{h}(x,t)+O(A^2), \quad q(x,t) = Q(\zeta) + A\hat{q}(x,t)+O(A^2), 
\end{equation}
where the behaviour of $\hat{h}$ and $\hat{q}$ is in some way related to the periodicity of the original travelling wave with respect to $\zeta$ with period $L$, and of the blowing and suction function with respect to $x$ with period $L_F$.

The
weighted-residual
equations for $h$ and $q$ yield, at $O(A)$,
\begin{equation}
\label{mass_trav}
 \hat{h}_t - f(x)+\hat{q}_x=0
\end{equation}
and
\begin{equation}
\label{flux_trav_wr}
 \hat{q} + w_6(\zeta) \hat{q}_t = w_0(\zeta) \hat{h} + w_1(\zeta) \hat{h}_x + w_2(\zeta) \hat{h}_{xxx} + w_3(\zeta) \hat{q} + w_4(\zeta) \hat{q}_x + w_5(\zeta)f(x),
\end{equation}
where
\begin{equation}
\label{w0_travelling}
 w_0 = H^2\left( 2- 2H' \cot\theta + \frac{H'''}{C}\right) -\frac{34 R Q Q'}{35},
\end{equation}
\begin{equation}
\label{w123_travelling}
 w_1 = -\frac{2}{3}H^3 \cot\theta + \frac{18R Q^2}{35}, \quad w_2 = \frac{H^3}{3C}, \quad w_3 = R\left( \frac{36 Q H'}{35} - \frac{34 H Q'}{35}\right),
\end{equation}
\begin{equation}
 w_4 = -\frac{34 R H Q}{35}, \quad w_5 = \frac{RHQ}{5}, \quad w_6 = \frac{2R H^2}{5}.
\end{equation}
The coefficients $w_i$, $i=0,\,...,\,6$ are functions of $\zeta$. If we set $f(x)=0$ in these equations, we recover the equations governing the evolution of linearised perturbations to the travelling wave, i.e. the equations of linear stability.
For non-zero $f$, we can integrate forward in time from a given initial condition
$\hat{h}(x,0)=\hat{h}_0(x)$, $\hat{q}(x,0)=\hat{q}_0(x)$,
 to determine $\hat{h}(x,t)$ and $\hat{q}(x,t)$. 

The system \eqref{mass_trav} and 
\eqref{flux_trav_wr}
features some terms that are known functions of $\zeta$, and others that are known functions of $x$, but the system is autonomous with respect to $t$. We therefore choose to rewrite the equations \eqref{mass_trav} and  
\eqref{flux_trav_wr}
in terms of $x$ and $\zeta$:
\begin{equation}
 \hat{h}_t \rightarrow -U \hat{h}_{\zeta}, \quad \hat{h}_x \rightarrow \hat{h}_x + \hat{h}_{\zeta},\quad  \hat{q}_t \rightarrow -U \hat{q}_{\zeta}, \quad \hat{q}_x \rightarrow \hat{q}_x + \hat{q}_{\zeta}.
\end{equation}
We obtain a system in two variables, $x$ and $\zeta$, with equations
\begin{equation}
\label{travelling_perturbation_1}
 -U \hat{h}_{\zeta} -f(x)+\hat{q}_x + \hat{q}_{\zeta}=0
\end{equation}
and
\begin{equation}
\begin{array}{ll}
\label{travelling_perturbation_2}
 \hat{q} - U w_6(\zeta) \hat{q}_{\zeta}  & = \quad   w_0(\zeta)\hat{h} + w_1(\zeta)(\hat{h}_x + \hat{h}_{\zeta}) + w_2(\zeta)(\hat{h}_{xxx} + 3\hat{h}_{x x \zeta} + 3 \hat{h}_{x \zeta \zeta } +\hat{h}_{\zeta \zeta \zeta} ) \\[8pt] & \quad \quad  + w_3(\zeta) \hat{q} + w_4(\zeta) (\hat{q}_x  + \hat{q}_{\zeta}) + w_5(\zeta) f(x).
 \end{array}
\end{equation}
We now regard $\zeta$ and $x$ as independent variables.
Under this transformation,
$x$ remains as a purely spatial variable, but $\zeta$ has a dual identity, incorporating both spatial and timelike components. 
The statement of initial conditions becomes more complicated in the new variables:
\begin{equation}
 \hat{h}(\zeta=x) = h_0(x), \quad \hat{q}(\zeta=x) = q_0(x)
\end{equation}
so that the initial conditions are spread across the whole range of $\zeta$.

If the Fourier expansion of $f(x)$ is
\begin{equation}
\label{F_fourier_series}
 f(x) = \sum_m f_m \cos{mx} + g_m \sin{mx},
\end{equation}
we can write the general solution of \eqref{travelling_perturbation_1} and \eqref{travelling_perturbation_2}
as
\begin{eqnarray}
\label{hq_fourier_series}
 \hat{h} = \sum_m J_m(\zeta)\cos{mx} + K_m(\zeta)\sin{mx} + h^*(\zeta, x),\\
 \hat{q} = \sum_m M_m(\zeta)\cos{mx}+N_m(\zeta)\sin{mx} + q^*(\zeta, x)
\end{eqnarray}
where the functions $J_m(\zeta)$, $K_m(\zeta)$, $M_m(\zeta)$ and $N_m(\zeta)$ are periodic in $\zeta$.
The remaining terms, $h^*$ and $q^*$, satisfy the homogeneous versions of \eqref{travelling_perturbation_1} and \eqref{travelling_perturbation_2} obtained by setting $f(x)=0$, which are exactly the equations for linear stability of the underlying travelling wave.
We can calculate the periodic functions $J_m$, $K_m$, ..., corresponding to a limit cycle, for any periodic travelling wave. However, we would only expect to observe the limit cycle behaviour in initial value calculations if the limit cycle is stable. If the travelling wave is stable, all solutions $h^*$, $q^*$ to the homogeneous problem eventually decay to zero and so $\hat{h}$ and $\hat{q}$ are limit cycles, periodic in time.

\begin{figure}
 \includegraphics{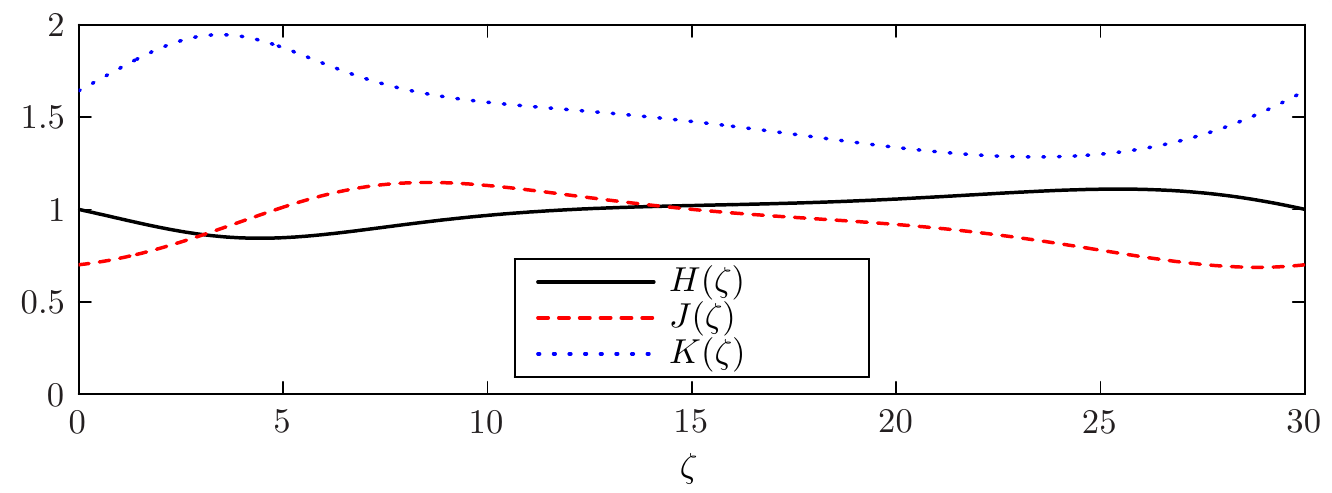}
\includegraphics{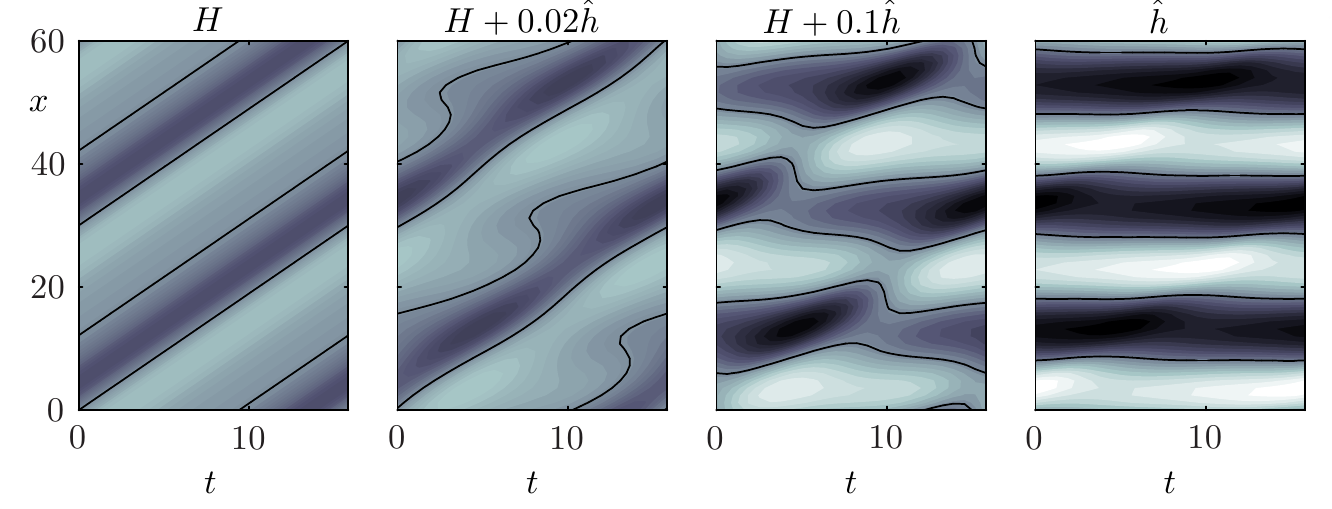}
\caption{Travelling wave $H(\zeta)$ and $O(A)$ perturbation $\hat{h}(x,t)$ for $\theta=\pi/4$, $C=0.05$, $R=3$, calculated with the weighted-residual equations. The initial travelling wave has wavelength $30$, but blowing and suction is applied with wavelength $20$;
we show solutions in a domain of length $60$. The black contours indicate $h=1$, and the same colour map is used in the first three figures.
Nonlinear time-dependent results for $A=0.02$ are shown in figure \ref{fig:Vary_ICs} for a selection of initial conditions.
\label{fig:Travelling_wave}
}
\end{figure}

We now calculate $\hat{h}$ for a small-amplitude limit cycle driven by $f(x)=\cos{mx}$, so that the forcing Fourier series has only a single component. The sums in \eqref{hq_fourier_series} are then over a single value of $m$, for which we obtain
\begin{equation}
 - U J_m' +mN_m + M_m'=1,
\end{equation}
\begin{equation}
 -U K_m' -mM_m + N_m'=0,
\end{equation}
\begin{equation}
\begin{array}{ll}
 M_m - w_6 U  M_m' & = \quad w_0 J_m + w_1 (J_m'+mK_m)+w_2(J_m''' + 3mK_m''-3m^2 J_m'-m^3 K_m) \\ [8pt]
 & \quad \quad + w_3 M_m + w_4(M_m'+mN_m)+
 w_5
 \end{array}
\end{equation}
and
\begin{equation}
\begin{array}{ll}
 N_m-w_6 U  N'_m & = \quad  w_0 K_m + w_1 (K_m'-mJ_m)+w_2(J_m''' + 3mK_m''-3m^2 J_m'-m^3 K_m)\\[8pt]
 & \quad \quad + w_3 N_m + w_4(N_m'-mM_m).
 \end{array}
\end{equation}
We solve the equations for $M = M_m$, $N =N_m$, $J = J_m$ and $K=K_m$ in \textsc{Auto-07p}, coupled to a system to find the travelling wave itself, seeking both travelling wave and perturbation periodic in $\zeta$ with period $L$.
Thus we obtain the solution
\begin{equation}
\label{h_hat_solution}
\hat{h} = J_m(\zeta) \cos{mx} + K_m(\zeta)\sin{mx} = J_m(x-Ut)\cos{mx} + K_m(x-Ut)\sin{mx}.
\end{equation}
Regardless of the relation between the travelling wave period $L$ and the blowing/suction wavenumber $m$, we see that if $J_m(\zeta)$ and $K_m(\zeta)$ are periodic with period $L$, then the function $\hat{h}(x,t)$ is temporally periodic with period $T=U/L$, which is the same temporal period as the base travelling wave. However, the solution for $\hat{h}(x,t)$ is spatially periodic only if $L/L_F$ is rational, with period given by the least common multiple of $L$ and $L_F$. 
A typical set of solutions for $H(\zeta)$, $J_m(\zeta)$ and $K_m(\zeta)$, and also the reconstructed field $\hat{h}$, are shown in figure \ref{fig:Travelling_wave}. For the travelling wave solution shown in figure \ref{fig:Travelling_wave}, we find that the functions $J_m(\zeta)$ and $K_m(\zeta)$ display little relative variation with $\zeta$. This means that the field $\hat{h}(x,t)$ is essentially steady.

\begin{figure}
 \includegraphics{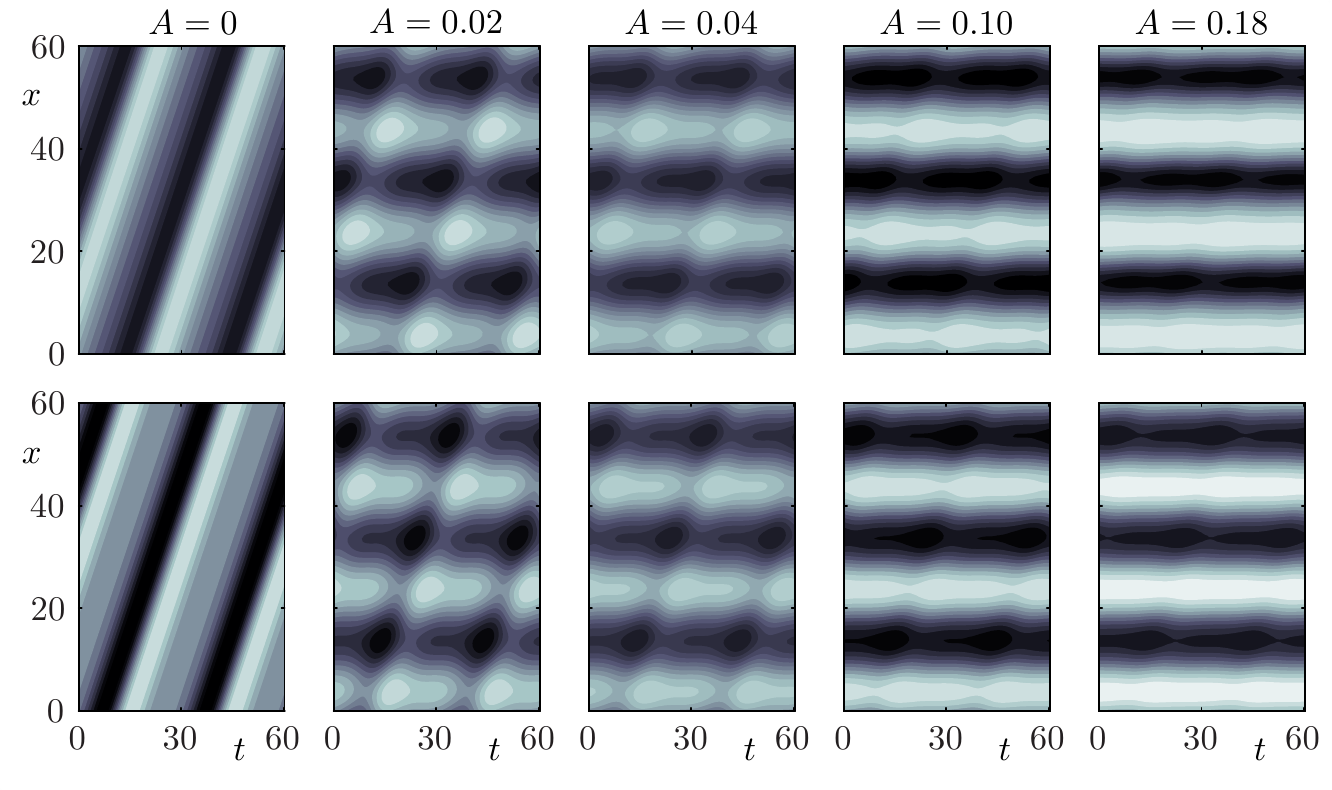}
 \caption{Contour plots of $h(x,t)$ from nonlinear time-dependent calculations (top row) and using the small-$A$ asymptotic solutions for perturbed travelling waves (bottom row) for $R=1.7$, $C=0.05$, $\theta=\pi/4$, in a domain of length $60$ with suction wavelength $L_F=20$. The colour map is scaled to the maximum and minimum value in each column.  Here the travelling wave is stable in the absence of suction. Time-dependent results for a larger $R$, for which the uniform film solution is unstable to multiple perturbations, are shown in figure~\ref{fig:Vary_ICs}.
\label{fig:IVP_and_perturbed_wave_R_1p7}
 }
\end{figure}

Figure \ref{fig:IVP_and_perturbed_wave_R_1p7} shows a comparison between fully nonlinear time-dependent calculations and the perturbed travelling wave solution at a value of $R$ small enough that the uniform state is unstable to only a single perturbation, and the travelling wave is stable. We obtain good agreement between the time-dependent calculations and the asymptotic predictions based on \eqref{Travelling_perturbation_expansion} and \eqref{h_hat_solution}. Both sets of results show a smooth transition as $A$ increases from travelling waves with wavelength $60$ at $A=0$ to almost-steady states at $A=0.18$ with wavelength $20$, which is wavelength of the blowing and suction.

\begin{figure}
 \includegraphics{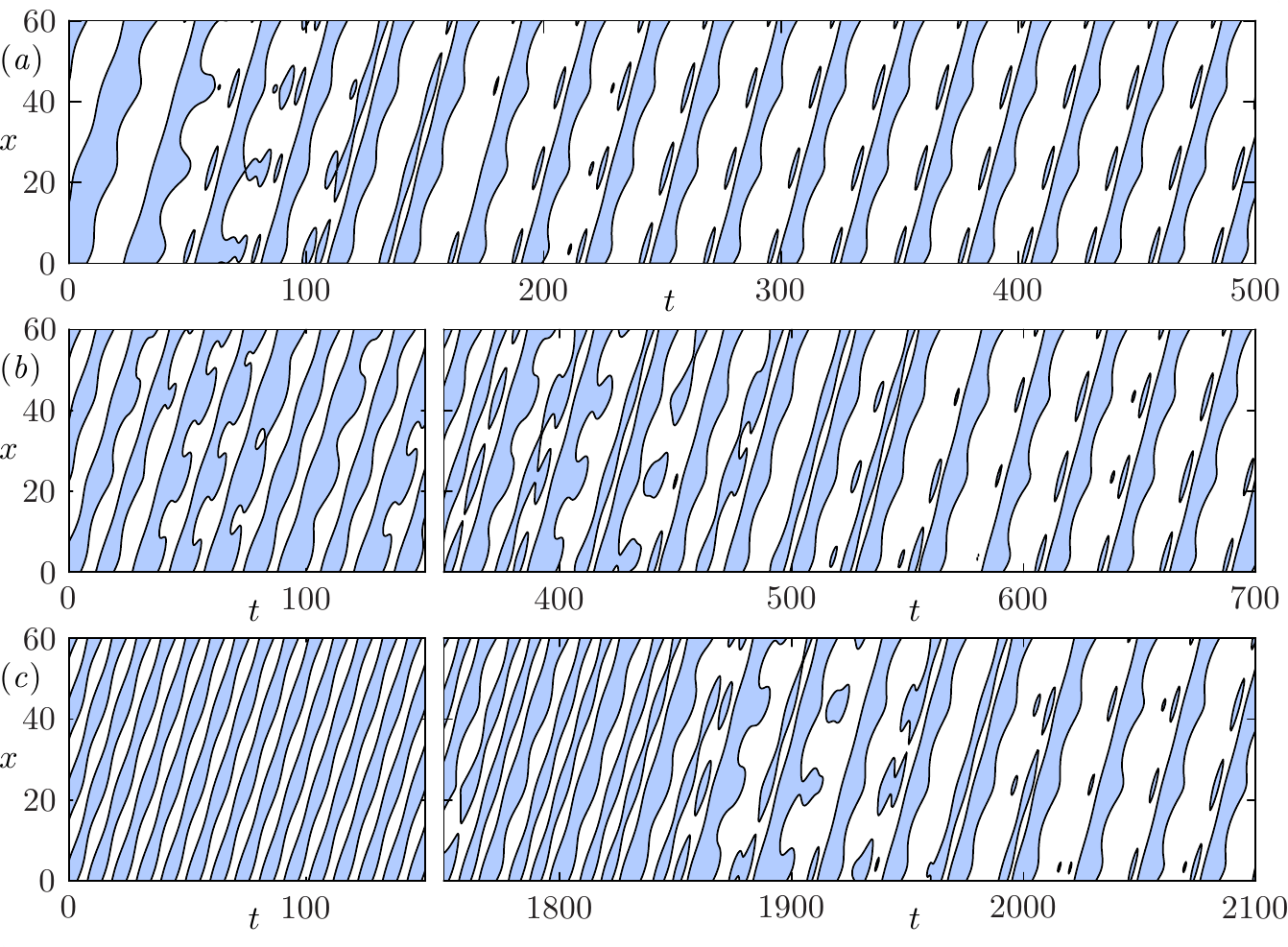}
 \caption{Time-dependent simulations in a domain of length 60, with  $F=0.02\cos(2\pi x/20)$, with $R=3$, $C=0.05$ and $\theta=\pi/4$, and varying initial conditions: we set $h(x,0)=1+0.1\cos(2\pi n x/60)$ and $q(x,0)=2/3 + 0.2\cos(2\pi n x/60)$, with $n=1,2,3$ in (\textit{a}-\textit{c}) respectively.
 These initial conditions correspond to neither travelling waves nor steady states, so nonlinear evolution is required if the system is to reach a periodic state. 
 These plots shows the single contour $h(x,t)=1$, with $h>1$ in the shaded region.
The same periodic state is reached eventually, regardless of the initial conditions.
 \label{fig:Vary_ICs}
 }
\end{figure}

The composite solution shown in figure \ref{fig:Travelling_wave} has $L_F=20$ and $L=30$; the resulting perturbed field has spatial period $60$. We have only considered corrections up to $O(A)$, and in this expansion, the imposed blowing and suction cannot affect the periodicity of the underlying travelling wave. However, in fully nonlinear time-dependent simulation, even if we begin with initial conditions that are perfectly periodic with period $L=30$, but force at a different wavelength, such as $L_F=20$, nonlinear effects will lead to a perturbation at wavelength $60$ that can cause the underlying travelling wave to double in spatial period, and likely change shape and speed. Eventually we would expect to reach the state where the dominant wave has period $60$, and the linear perturbation field $\hat{h}$ is a solution to equations where the coefficients
$w_i$
are those for the travelling wave with wavelength 60. 
Figure \ref{fig:Vary_ICs} shows time-dependent simulations for forcing at wavelength $20$ with $A=0.02$ in a domain of length $60$, at a value of $R$ large enough that travelling waves with wavelengths $60$, $30$ and $20$ are unstable.
When the initial conditions involve only modes of wavelength 60, a periodic initial condition is quickly reached. For initial conditions of wavelength 30, these modes compete with the wavelength 20 forcing, but eventually a wavelength 60 state is indeed achieved. For initial conditions with wavelength equal to the initial forcing, we rapidly reach an initial periodic state, with three equal waves in the domain. However, after a very long time, noise in the system causes a switch to a single wave with wavelength $60$, thus yielding the same state regardless of initial conditions.

\section{Initial value problems}
\label{sec:IVP}

Even in the absence of blowing and suction, a thin liquid film falling along an inclined plane can display rich behaviour, including pattern formation, transition to chaos, travelling waves, and pulse-like structures.
Many of the instabilities that dominate the dynamics occur at long wavelength, and so the system is frequently studied using long-wave models. However, a well-known feature of the Benney model is that the interface can exhibit finite-time flow up at Reynolds numbers larger than critical \citep{Pumir, Ruyer_Quil}, which is not replicated in full Navier--Stokes simulations. The weighted-residual equations were developed partly to avoid this blow-up behaviour, and are indeed better behaved that the Benney models. The two models agree when the system parameters are close to neutral stability.

The introduction of forcing in the form of suction boundary conditions introduces another potential mechanism for finite-time blow-up, whereby the film thins to zero in finite time, as illustrated in figure \ref{fig:Subharmonic_crash}. This thinning occurs even at zero Reynolds number, and so arises in both Benney and weighted-residual models. The blow-up need not occur at the same wavelength as the forcing function; for example it can be triggered by subharmonic-perturbations to a periodic steady state. 

Blow-up is generally avoided if $R$ is not too large, and blowing and suction is applied with sufficiently small amplitude, but the film dynamics can still interact with the imposed suction. In the absence of suction, the system exhibits periodic behaviour in the form of travelling waves, which propagate at a constant speed without changing form. If non-uniform suction is imposed via a non-constant function $F(x)$, which remains fixed with respect to the frame of the wall, 
waves must change in form as they propagate, 
and so time-periodic behaviour manifests as nonlinear limit cycles which depend on both $x$ and $t$. Figure \ref{fig:IVP_and_perturbed_wave_R_1p7} shows initial value calculations in which the unforced system exhibits stable travelling waves. The imposed suction need not have the same wavelength as the travelling wave, and in figure \ref{fig:IVP_and_perturbed_wave_R_1p7}, three periods of suction fit within the discretised domain. The initial value calculations show a smooth transition as $A$ increases from travelling waves at $A=0$ to an essentially steady state at $A=0.18$. At small $A$, the blowing and suction causes slight perturbations to the travelling wave, while at larger $A$, small-amplitude disturbances propagate over a large-amplitude non-uniform steady state. The underlying flow field retains its dominant down-slope direction, and so there is still a non-zero perturbation wave speed even when the interface shape is steady.

\begin{figure}
\subfloat[$A=0.25$, $R=4$]{\includegraphics{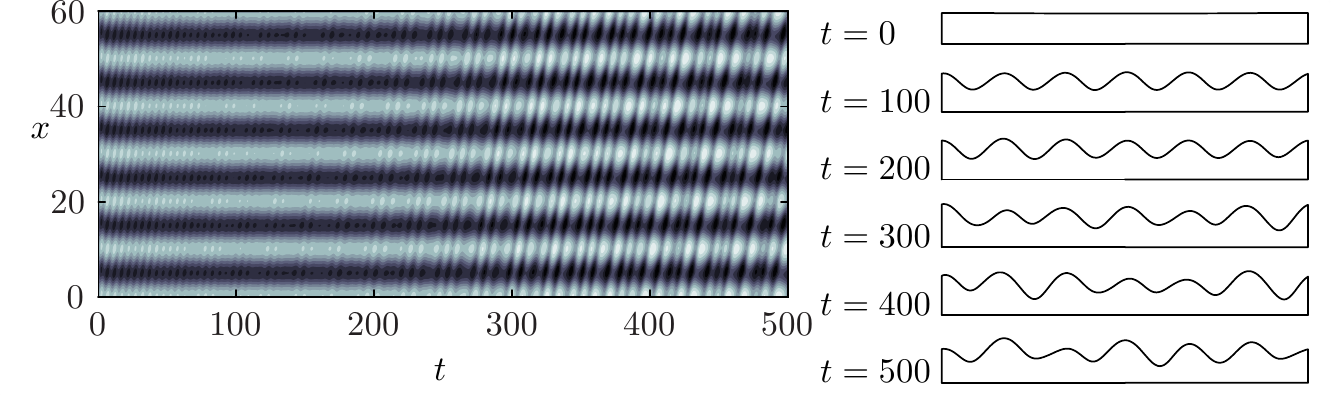}}

\subfloat[$A=0.45$, $R=4$]{\includegraphics{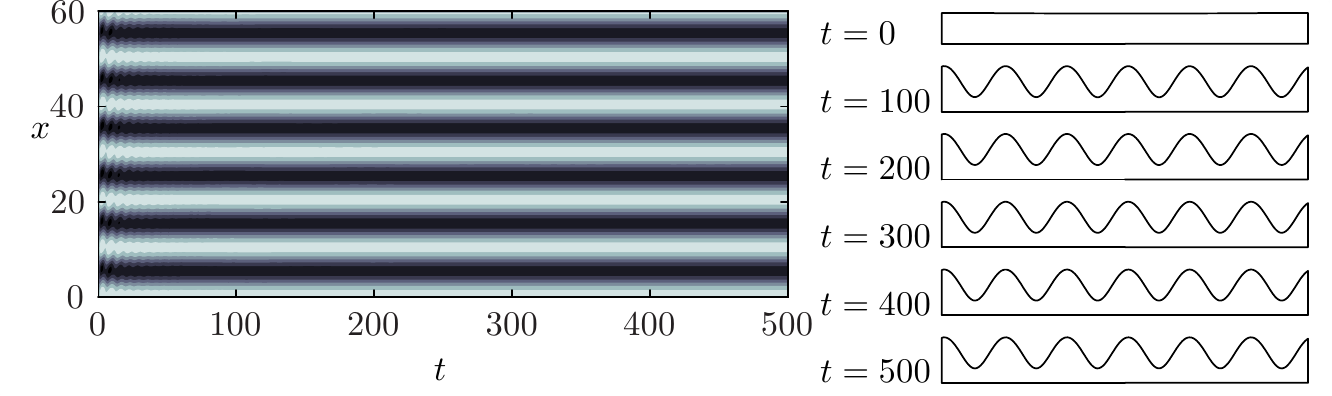}}

\subfloat[$A=0.45$, $R=2$]{\includegraphics{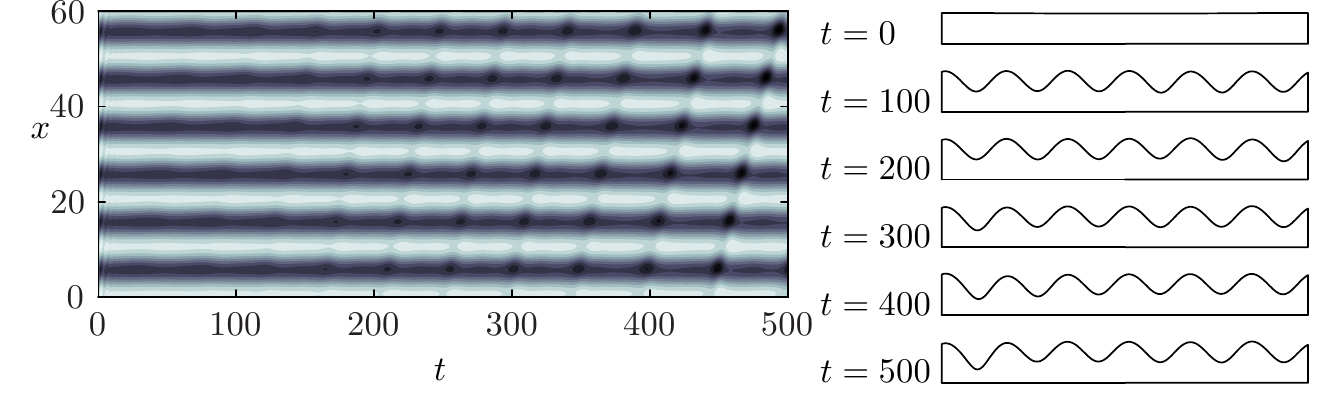}}
\caption{Nonlinear time-dependent calculations for $L=10$, $C=0.05$, $\theta=\pi/4$. The simulations are conducted in a domain of length $6L$, and the initial conditions for $h$ and $q$ includes a small perturbation proportional to $\sin(2\pi x/(6L))$.
The stable solution shown in (\textit{b}) can be accessed by either increasing the Reynolds number from (\textit{c}), or the amplitude of blowing and suction from (\textit{a}).
\label{fig:stable_region}
}
\end{figure}

The applied suction can also lead to states which are neither steady nor limit cycles. Figure \ref{fig:stable_region} shows the result of initial value calculations conducted at parameters near to the stable `island' shown in figure \ref{fig:RA_maps} for $L=10$. This is a region of parameter space with steady solutions that are stable to perturbations of all wavelengths, but is unusual in that solutions lose stability if either $A$ or $R$ is decreased, so that sufficiently large inertia is required to maintain stability.
 In case shown in figure \ref{fig:stable_region}(\textit{a}), with $A=0.25$ and $R=4$, the height field displays aperiodic behaviour. At each instant, six peaks corresponding to the six periods of the suction function are visible, but waves propagate over these peaks without a clear structure.
 We can increase the suction amplitude to $A=0.45$ to obtain steady, stable solution, shown in figure \ref{fig:stable_region}(\textit{b}).
Decreasing the Reynolds number to $R=2$ means that the steady state again loses stability, but in this case to a wavelength $60$ travelling wave which propagates over the steady solution in a regular manner, as shown in figure \ref{fig:stable_region}(\textit{c}).
 
\section{Conclusion}
\label{sec:Conclusion}

In this paper, we considered thin-film flow down an inclined plane modified so that fluid is injected and withdrawn through the wall according to an arbitrary function $F(x,t)$. 
We derived and studied two long-wave models, based on the first-order Benney and weighted-residual formulations, for this system.
If the average layer height is conserved, then $F$ must have zero mean in space. We then specialised to the case where $F$ is a single steady Fourier mode, $F=A\cos{mx}$, and investigated the form, bifurcations and linear stability of steady states, as well as a range of fully-nonlinear time-dependent behaviours.

Any steady states subject to non-uniform blowing and suction $F$ must themselves be non-uniform. We calculated the interface shape for small $A$, and showed that this differs between the two models when $R\neq 0$, but that the results agree in the long-wave limit. There is a phase difference between the shape of the blowing and suction and interface height, which originates from the preferred down-slope direction of the base flow.

As $A$ increases, the interface becomes increasingly non-uniform, and the nonlinear terms in the governing equations become increasingly important. Two notable transitions occur as $A$ increases: firstly all steady states feature regions with negative volume-flux $q$, and secondly steady solutions disappear altogether for sufficiently large $A$. Time-dependent evolution beyond the existence of steady states shows that the film can dry in finite time, as the fixed rate of fluid removal via $F$ is not matched by the supply of fluid.

We analysed the behaviour of the streamfunction in order to interpret steady solutions with negative $q$. We found that imposing non-zero, non-uniform $F$ leads to a series of isolated stagnation points along the wall, wherever $F(x)=0$. These stagnation points are either saddle points or local extrema in the streamfunction, depending on the sign of $F'(x)$ and $q(x,t)$ at these points. Connections between the stagnation points allow us to demarcate the boundary between fluid which enters and leaves through the wall, and `propagating' fluid which never reaches the wall.
At $A=0$, all fluid propagates down the wall, but the height of the recirculating layer near the wall increases with $A$.
For large enough $A$, all steady solutions have negative $q$ for some $x$. For those $x$ with $q(x)<0$, all flow is directed up the slope, against gravity; this flow reversal means that the propagating layer disappears.

Steady solutions need not have the same spatial periods as the imposed suction, and
subharmonic steady solutions can emerge via pitchfork bifurcations from the harmonic solution branches. In our calculations, we have only found subcritical pitchfork bifurcations, leading to unstable subharmonic steady states.
Unstable subharmonic steady states also occur in thin-film flow with periodic topography \citep{Inclined_topography}, but it is possible that subharmonic steady states may be stable for other parameter values.
 Initial-value calculations starting from near the periodic steady states sometimes show subharmonic drying, where the film dries at any one of the local minima of $h$.

In the absence of suction, the primary instability of the uniform state is to eigenfunctions with complex eigenvalues, leading to Hopf bifurcations which do not appear in the bifurcation structure for steady solutions.
If $F$ is spatially periodic, then the steady states are also periodic, as are the coefficients of the linear equation governing the evolution of small perturbations. This means that perturbations must propagate through a heterogeneous environment, and also that the eigenmodes are no longer simply exponentials of the form $\exp(ikx+\lambda t)$. Instead, according to the Floquet-Bloch form, the eigenmodes can be written as $g(x) \exp(ikx+\lambda t)$, where $g(x)$ is a complex-valued function, spatially periodic with the same period as the base state.
We calculated linear stability numerically for two classes of perturbations: firstly those that are periodic with the same period as the blowing and suction, and secondly perturbations of any wavelength. We found that introducing spatially-periodic suction with amplitude $A$ can increase or decrease the critical Reynolds number for linear stability to perturbations of both classes.
Due to the symmetries of the system, the correction to the eigenvalue, and hence to the critical Reynolds number, is even in $A$.

For small $A$, we calculated the critical Reynolds number to $O(A^2)$ for perturbations of arbitrary wavenumber. Imposing blowing and suction at very long wavelength increases the critical Reynolds number for both classes of perturbations when $\theta<\pi/2$. 
Due to the quadratic dependence of the critical $R$ on perturbation wavenumber, the first eigenmodes to become unstable as $R$ increases are those with infinite wavelength, both when $A=0$ and for small $A$, 
In contrast, for the case of flow underneath an inclined plane, where $\theta>\pi/2$, small-amplitude long-wave blowing and suction always reduces the critical Reynolds number, and so has a destabilising effect on the flow. 
However, as the small $A$ expansion is valid only close to the critical Reynolds number, which is negative for $\theta>\pi/2$,
the predicted dependence of stability properties on $A$ may not hold for physically realisable flows.

Determination of the largest $R$ that can be stabilised in an infinite domain by imposing steady suction is difficult as it is not clear that long-wavelength waves are always the most unstable, and both the steady solutions and their corresponding linear stability operators must be computed at finite $A$.
We find that the weighted-residual model predicts large increases in the critical Reynolds number for stability to perturbations of all wavelengths, from $1.25$ to $4$ or more, if blowing and suction is introduced with wavelengths of around 200. However, the imposed suction appears to become less effective at stabilising the flow if applied with very long wavelength.

In the absence of suction, the system can support travelling waves which propagate at a constant speed $U$ without changing form. Periodic travelling waves are periodic with respect to both space and time, but their dependence on $x$ and $t$ can be expressed in terms of a single variable $\zeta = x-Ut$. Imposing blowing and suction as $F(x)$ introduces heterogeneity to the system, and so we expect travelling waves to become limit cycles that are periodic in both $x$ and $t$, with explicit dependence on both variables. We calculated the $O(A)$ corrections to the travelling wave, and showed that the equations are closely related to those governing linear stability of the travelling wave. We demonstrated that limit cycles at $O(A)$ can be decomposed by rewriting the equations in terms of the independent variables $x$ and $\zeta$. The resulting solution for $h(x,t)$ and $q(x,t)$ is periodic with respect to time with the same time period as the underlying travelling wave, but is spatially periodic only if the ratio between the wavelength  of the blowing and suction and the wavelength of the travelling wave is rational.
We showed that the predictions of the $O(A)$ perturbation analysis are in good agreement with the results of fully nonlinear simulations.
 We also conducted initial-value calculations to investigate competition between the period of the travelling wave and of the imposed suction. We found that the same state is eventually reached, but this competition can persist over a large number of cycles.

We observed in \S~\ref{sec:Form_of_forcing} that the derivation of our equations is unchanged if the blowing and suction profile additionally depends on time with sufficiently long timescale.
Allowing time-dependence of the blowing and suction leads to the possibility of using $F$ to explicitly control the flow, and thus to stabilise otherwise unstable states by acting in response to the development of perturbations. For example, with perfect implicit knowledge of the system, we could choose $F(x,t)$ in order to drive the system towards a particular steady state.
However, the additional independent degree of freedom that emerges in the weighted-residual model may be important when exploring control of the system, as we cannot choose $F(x,t)$ to simultaneously specify $h(x,t)$ and $q(x,t)$.
Under the assumption of small deviations from a uniform state, with small $F$, both the weighted-residual equations and the Benney equations reduce via a weakly nonlinear analysis to a forced version of the Kuramoto--Sivanshinsky equations, for which optimal controls have been successfully applied \citep{Susana}.
Work is in progress \citep{Control_paper} to explore the effectiveness of feedback control strategies based on the nonlinear long-wave models derived in this paper.

\section*{Acknowledgements}

This work was funded through the EPSRC grant EP/K041134/1.

\bibliographystyle{jfm}

\bibliography{Bibliography}

\appendix

\section{Derivation of long-wave equations}
\label{sec:Derivation_appendix}

We first rescale the governing equations \eqref{non-scaled-start} to \eqref{non-scaled-end} according to 
\begin{equation}
 X = \delta x, \quad T = \delta t, \quad v = \delta w, \quad C = \delta^2 \widehat{C},
\end{equation}
where we are interested in the long-wave limit $\delta \ll 1$. 
We will consider two different scalings for $F$: firstly $F = \delta f$, and also $F = \delta^2 \hat{f}$. If $F$ is unsteady, we require for consistency that $f_T =O(1)$ or smaller.
The full set of equations becomes
 \begin{eqnarray}
 \label{scaled-start}
R\delta\left(u_T  + u u_X +  w u_y\right) 
&=& -  \delta p_X
 +2+\delta^2u_{XX}+ u_{yy},\\
R \delta^2\left(w_T + u w_{X}   + w w_y \right) &=& 
-  p_y
-2 \cot\theta 
+\delta^3 w_{XX}+ \delta w_{yy},\\
\label{appendix_mass_conservation}
u_X + w_y&=&0.
\end{eqnarray}
The boundary conditions at $y=0$ are
\begin{eqnarray}
\label{appendix_wall_bcs}
 u=0, \quad w=f(X,T),
\end{eqnarray}
and at $y=h$ we have
\begin{eqnarray}
\left (\delta^2 w_X+u_y\right)\left (1-\delta^2 h_X^2\right) + 2\delta h_X \left(\delta w_y-\delta u_X\right)&=&0,\\
 p - p_a
 - \frac{2}{1+ \delta^2 h_X^2} 
 \left(
 \delta w_y
+ \delta^3 u_X h_X^3  - \delta h_X\left(\delta^2 w_X+u_y\right)\right) 
&=& -\frac{h_{XX}}
 {
 \widehat{C}(1+\delta^2 h_X^2)^{3/2}
 }.\,\,\,
 \label{scaled-end}
\end{eqnarray}
The rescaled kinematic equation \eqref{mass_governing}
is 
\begin{equation}
\label{mass_longwave}
 h_T - f(X,T) + q_X=0, \quad q = \int_0^h u \, \mathrm{d}y.
\end{equation}
As $f$ is known, we need only an expression for $q$ to close the system. 

\subsection{Benney equation}
\label{sec:Benney_derivation}

To obtain the Benney equation for $q$ as a function of $h$, we expand $u$, $w$, $p$ and $q$ in powers of $\delta$, while assuming that $h$ is an $O(1)$ quantity:
\begin{equation}
 u = u_0 + \delta u_1 +O(\delta^2), \quad  w = w_0 + \delta w_1 +O(\delta^2)
\end{equation}
\begin{equation}
 p = p_0 + \delta p_1 +O(\delta^2),\quad  q = q_0 + \delta q_1 +O(\delta^2) . 
\end{equation}
The leading-order solution of \eqref{scaled-start} to \eqref{scaled-end} for small $\delta$ is
\begin{equation}
 u_0 = y(2h-y), \quad w_0 = f(X,T) - y^2 h_X, \quad p_0 = p_a - \frac{h_{XX}}{\widehat{C}} +2(h-y)\cot{\theta}.
\end{equation}
The flux at this order is
\begin{eqnarray}
 q_0 = \int_0^h u_0 \, \mathrm{d}y = \frac{2h^3}{3},
\end{eqnarray}
leading to the evolution equation
\begin{eqnarray}
\label{Benney_leading}
 h_T - f(X,T) + 2h^2 h_X= O(\delta).
\end{eqnarray}
We must calculate $u_1$ in order to obtain the $O(\delta)$ correction to $q$.
After integrating the $O(\delta)$ part of \eqref{scaled-start} twice, and applying boundary conditions at this order, we find 
\begin{equation}
  u_{1} = \frac{y \hat{p}_X }{2}(y-2h)+R\left[
(h_T-f) \left(\frac{y^3}{3} - h^2 y\right)  
+ \frac{2h h_X }{3}\left(\frac{y^4}{4} - h^3 y\right)  
+  h y (y-2h)f \right],
\end{equation}
where we have defined
\begin{eqnarray}
 \hat{p} = 2 h\cot\theta - \frac{h_{XX}}{\widehat{C}}.
\end{eqnarray}
The first-order correction to the flux can now be calculated as
 \begin{eqnarray}
 \label{q1-with-t}
 q_{1} =\int_0^h u_1 \, \mathrm{d}y=
 -\frac{h^3 \hat{p}_X }{3}
+
R\left(
-\frac{5h_T h^4}{12} 
-\frac{3h^6h_X }{10}
- \frac{h^4 f}{4}
\right).
\end{eqnarray}
We eliminate $h_T$ from this expression by using \eqref{Benney_leading},
and thus obtain the Benney equation for $q$:
\begin{equation}
\label{Benney_q_with_delta}
 q = q_0+ \delta q_1 + O(\delta^2)= 
 \frac{2h^3}{3} 
 -\delta\frac{h^3 \hat{p}_X }{3}
+
R\delta \left( 
\frac{8h^6h_X }{15}
-\frac{ 2h^4 f(X,T)}{3}
\right) + O(\delta^2).
\end{equation}

Alternatively, if $f=\delta \hat{f}$, then $\hat{f}$ still appears via the mass conservation \eqref{mass_longwave}, but is absorbed into the $O(\delta^2)$ error term in \eqref{Benney_q_with_delta}.

\subsection{First-order weighted-residual equations}
\label{sec:WR_derivation}

The derivation of the first-order weighted-residual equations closely follows the original derivation presented by \citet{Ruyer_Quil}. Imposing suction at the wall does not affect the boundary conditions on $u$, and so the basis functions for $u$ described by \citeauthor{Ruyer_Quil} yield the first-order equation without difficulty.

In contrast to the derivation of the Benney equations, here we use $\delta$ as an ordering parameter, rather than directly expanding variables with respect to $\delta$. For the first-order weighted-residual equations, we retain terms up to and including $O(\delta)$ in the equations, and so the momentum equations yield
 \begin{eqnarray}
 \label{eq:x_momentum_WR}
R\left(\delta u_T  + \delta u u_X + \delta w u_y\right) 
= -  \delta p_X
 +2+ u_{yy} +O(\delta^2)
\end{eqnarray}
and
\begin{eqnarray}
\label{eq:y_momentum_WR}
0= 
-  p_y
-2 \cot\theta + \delta w_{yy} +O(\delta^2),
\end{eqnarray}
while the mass conservation equation \eqref{appendix_mass_conservation} and the boundary conditions on the wall \eqref{appendix_wall_bcs} are unchanged.
The normal and tangential components of the dynamic boundary condition become
\begin{eqnarray}
\label{eq:dynamic_bc_WR}
0=u_y+O(\delta^2) , \quad 
 p
 = p_a + 2\delta u_X - \frac{h_{XX}}
 {
 \widehat{C}
 }+O(\delta^2) \quad \text{at $y=h$.}
\end{eqnarray}

We can integrate \eqref{eq:y_momentum_WR} with respect to $y$, and apply \eqref{eq:dynamic_bc_WR} to obtain
\begin{equation}
\label{p_integral_WR}
 p = p_a+ 2 (h-y)\cot\theta - \frac{h_{XX}}{\widehat{C}}
  -\delta u_X - 2\delta h_X u_y .
\end{equation}
We then substitute \eqref{p_integral_WR} into \eqref{eq:x_momentum_WR} and discard terms smaller than $O(\delta)$, leaving
 \begin{eqnarray}
 \label{eq:WR_xmomentum_no_p}
R\delta\left( u_T  +  u u_X + w u_y\right) 
= -\delta \hat{p}_X
 +2+ u_{yy},
\end{eqnarray}
which is coupled to the mass conservation equation \eqref{appendix_mass_conservation},
and subject to the boundary conditions $u=0$ and $w=f(X,T)$ at $y=0$, and $u_y=0$ at $y=h(X,T)$.

Following \citeauthor{Ruyer_Quil}, we posit an expansion for $u$ in terms of basis functions $\phi_j$ satisfying no-slip on the wall and zero tangential stress on the interface:
\begin{equation}
 u = \sum_{j} a_j(X,T) \phi_j(\bar{y}), \quad
 \phi_j(z) = z^{j+1}-\left(\frac{j+1}{j+2}\right)z^{j+2}, \quad \bar{y} = \frac{y}{h(X,T)}.
\end{equation}
If $\delta=0$, the only non-zero $a_n$ is $a_0$, and for small $\delta$, we find that $a_0=O(1)$, while $a_n=O(\delta)$ or smaller for $n\geq 1$.
We can calculate the coefficients $a_j$ by using the $\phi_j$ as test functions in \eqref{eq:WR_xmomentum_no_p}, but can obtain the leading order equations by use of $\phi_0$ only.

Correct to $O(\delta)$ we can write the weak form of \eqref{eq:WR_xmomentum_no_p} as
\begin{equation}
\label{WR_ordered}
 R \delta \int_0^h \phi_n(\bar{y}) \left( u_{0T}  +  u_0 u_{0X} + w_0 u_{0y}\right) \,\mathrm{d} y
 =
 \left(-\delta\hat{p}_X+2\right) \int_0^h \phi_n(\bar{y}) \,\mathrm{d} y
 + \int_0^h \phi_n(\bar{y}) u_{yy}\,\mathrm{d} y,
\end{equation}
where
\begin{equation}
 u_0(X,y,T) = \frac{3q}{h}\phi_0(\bar{y}), \quad w_0(X,y,T) = f(X,T) - \int_0^y u_{0X}(X,y', T)\, \mathrm{d}y'.
\end{equation}

After setting $n=0$ and repeated integration by parts, \eqref{WR_ordered} yields
\begin{eqnarray}
\label{WR_q_with_hT}
R \delta \left( \frac{2}{5} q_T - \frac{23}{40} \frac{q h_T}{h}+\frac{111}{280} \frac{q q_X}{h}
-\frac{18}{35}\frac{q^2 h_X}{h^2}
+\frac{3qf}{8h}
\right) =\left(-\delta \hat{p}_X + 2\right)\frac{h}{3} -\frac{q}{h^2}.
\end{eqnarray}
We can eliminate $h_T$ from \eqref{WR_q_with_hT} by using \eqref{mass_longwave}, and thus obtain
\begin{equation}
\label{WR_q_with_delta}
 q + \frac{2}{5} R \delta h^2 q_T = \frac{2h^3}{3} - \frac{\delta h^3 \hat{p}_X}{3} + R \delta \left( \frac{18}{35} q^2 h_X - \frac{34}{35}h q q_X + \frac{hqf(X,T)}{5}\right) + O(\delta^2).
\end{equation}
The two equations \eqref{mass_longwave} and \eqref{WR_q_with_delta} form a closed system for $h$ and $q$ without requiring calculation of $a_n$ for $n\geq 1$.
However, the higher coefficients are required to fully determine the velocity field within the fluid layer.

As in the case of the Benney equations, if we take $f = \delta\hat{f}$, the mass conservation equation \eqref{mass_longwave} is unchanged, but the term involving $f$ in \eqref{WR_q_with_delta} disappears as it is absorbed into the $O(\delta^2)$ error term.

\end{document}